\documentclass[12pt,reprint,aps,groupedaddress,nofootinbib,prd,twocolumn]{revtex4-2}

\usepackage[english]{babel}
\usepackage[utf8]{inputenc}
\usepackage{amsmath,amssymb,graphicx,amsfonts}

\usepackage{epsfig}
\usepackage[colorlinks=true,
linkcolor=blue,
urlcolor=blue,
citecolor=blue]{hyperref}
\usepackage{bm}
\usepackage{mathrsfs}
\usepackage{enumerate}
\usepackage{amsthm}
\usepackage{bbm}
\usepackage{comment}
\usepackage{physics}
\usepackage{url}
\usepackage[dvipsnames]{xcolor}

\usepackage[title]{appendix}

\font\tenscr=rsfs10 scaled1100
\font\sevenscr=rsfs7 
\font\fivescr=rsfs5 
\skewchar\tenscr='177
\skewchar\sevenscr='177
\skewchar\fivescr='177
\newfam\scrfam
\textfont\scrfam=\tenscr
\scriptfont\scrfam=\sevenscr
\scriptscriptfont\scrfam=\fivescr

\def\scri{{\fam\scrfam I}}

\newcommand{\nn}{\nonumber}

\begin{document}

\title{Pseudospectrum of horizonless compact objects: a bootstrap instability mechanism}

\author{Valentin Boyanov$^{1,2}$, Kyriakos Destounis$^{3}$, Rodrigo Panosso Macedo$^{4}$, Vitor Cardoso$^{2,5}$, Jos\'e Luis Jaramillo$^{6}$} 
\affiliation{$^1$Departamento de F\'{\i}sica T\'eorica and IPARCOS, Universidad Complutense de Madrid, 28040 Madrid, Spain}
\affiliation{${^2}$CENTRA, Departamento de F\'{\i}sica, Instituto Superior T\'ecnico -- IST, Universidade de Lisboa -- UL, Avenida Rovisco Pais 1, 1049 Lisboa, Portugal}
\affiliation{$^3$Theoretical Astrophysics, IAAT, University of T{\"u}bingen, 72076 T{\"u}bingen, Germany}
\affiliation{$^4$STAG Research Centre, University of Southampton, University Road SO17 1BJ, Southampton, UK}
\affiliation{${^5}$ Niels Bohr International Academy, Niels Bohr Institute, Blegdamsvej 17, 2100 Copenhagen, Denmark}
\affiliation{$^6$Institut de Math\'ematiques de Bourgogne (IMB), UMR 5584, CNRS, Universit\'e de Bourgogne Franche-Comt\'e, F-21000 Dijon, France}

\begin{abstract}
Recent investigations of the pseudospectrum in black hole spacetimes have shown that quasinormal mode frequencies suffer from spectral instabilities. This phenomenon may severely affect gravitational-wave spectroscopy and limit precision tests of general relativity. We extend the pseudospectrum analysis to horizonless exotic compact objects which possess a reflective surface arbitrarily close to the Schwarzschild radius, and find that their quasinormal modes also suffer from an overall spectral instability. Even though all the modes themselves decay monotonically, the pseudospectrum contours of equal resonance magnitude around the fundamental mode and the lowest overtones can cross the real axis into the unstable regime of the complex plane, unveiling the existence of non-modal pseudo-resonances. A pseudospectrum analysis further predicts that fluctuations to the system may destabilize the object when next to leading-order effects are considered, as the triggering of pseudo-resonant growth can break the order-expansion of black-hole perturbation theory.
\end{abstract}

\maketitle

\section{Introduction}

The historic detection of gravitational waves (GWs) \cite{LIGOScientific:2020ibl} paved the way for precision GW astrophysics and black hole (BH) spectroscopy to develop at an unprecedented rate. The increase in sensitivity of ground-based interferometers, as well as the arrival of next generation space-borne detectors, will further improve our understanding of the gravitational interaction \cite{Barack:2018yly,Barausse:2020rsu,Amaro-Seoane:2022rxf}. BH and compact object binaries produce GWs through the coherent motion of the sources and form ideal testbeds of the gravitational field in extreme conditions. Together with new radio and deep infrared interferometry \cite{Doeleman:2008qh}, which provide direct images of the dark compact objects lurking at galactic centers, we are now able to perform novel tests of issues such as the BH no-hair theorem \cite{Cardoso:2016ryw}, dark matter and environmental effects \cite{Macedo:2013qea,Annulli:2020ilw,Annulli:2020lyc,Vicente:2022ivh,Barausse:2014tra,Cardoso:2021wlq}, potential quantum effects at the horizon scale \cite{Cardoso:2016rao,Conklin2021} as well as the existence of horizonless exotic compact objects (ECOs) \cite{Macedo:2013jja,Cardoso:2016rao,Giudice:2016zpa,CardosoPani2019}.

Our current understanding of stellar evolution suggests that when astrophysical objects become sufficiently compact, the interactions within their matter content can no longer counteract their enormous self-gravity. The natural conclusion to their evolution would therefore be a collapse into a BH. Nevertheless, the singularity present at the end of this classical collapse process suggests that this picture is incomplete. Corrections may, on the one hand, be localized to a small region around the singularity itself, leaving the exterior as an effectively classical BH. On the other hand, they may permeate outwards up to the horizon, or even slightly past it, removing the classical trapped region altogether. This line of thought has inspired the search for BH mimickers in the form of effectively classical horizonless ECOs, such as boson stars \cite{Liebling:2012fv}, gravastars \cite{Mazur:2001fv}, or semiclassically sustained ultracompact objects \cite{Carballo-Rubio:2017tlh,Arrechea:2021pvg}, as well as other modifications with exotic near-horizon structure such as firewalls \cite{Kaplan:2018dqx} and fuzzballs \cite{Mathur:2005zp,Mathur:2008nj,Ikeda:2021uvc}.

It is generally accepted that the ringdown signal of a Kerr BH will be dominated by its quasinormal modes (QNMs) \cite{Kokkotas:1999bd,Berti:2009kk}, with the QNM spectrum being characterized by the BH's mass and spin. This is due to the fact that the ringdown frequencies coincide (see e.g. \cite{zworski2017mathematical,dyatlov2019mathematical} and references therein) with the QNM frequencies (which describe excitations of the photon sphere). However, it has been shown \cite{Cardoso:2016rao,Cardoso:2016oxy} that the relaxation of a highly-compact horizonless object is almost identical to that of a BH at early times, consisting of the usual photon sphere modes \cite{Cardoso:2008bp,Cardoso:2017soq,Cardoso:2018nvb,Destounis:2018qnb}, even though these no longer correspond to QNMs. The replacement of the event horizon by a surface with at least partial reflectivity entraps modes between it and the photon sphere, which subsequently emerge during the later stages of the ringdown and manifest as successively damped repetitions of the initial signal (known as echoes~\cite{Price:2017cjr}) which describe the proper vibrational spectrum of the ECO itself. The QNMs of BHs and ECOs can differ dramatically since they are defined by eigenvalue problems with different boundary conditions. Particularly, the fundamental mode and first overtones of non-rotating ECOs with high compactness become extremely long-lived (with very long decay timescales) \cite{Cardoso:2016rao,Barausse:2014tra}.

These phenomena are not restricted to ECOs, but are commonly present in more realistic compact objects. It is already well established that long-lived QNMs are present in compact neutron stars~\cite{Kokkotas1994,Chandrasekhar_1991}. These modes manifest themselves in waveforms as a repeated resurgence of the initial part of the signal, typically referred to as ``echoes"~\cite{Kokkotas:1995av,Tominaga:1999iy,Ferrari:2000sr}, although a precise interpretation in terms of light ring and cavity excitation was only put forward more recently~\cite{Cardoso:2019rvt}. In the case of BH mimickers, however, these long-lived modes may well turn out to destabilize the object when effects beyond linear-order perturbations are taken into consideration \cite{Cardoso:2014sna,Keir:2014oka,Cunha:2022gde}. This becomes more clear when rotation is introduced; ultracompact objects with surfaces within the ergosphere suffer from the ergoregion instability \cite{Friedman78,Schutz78} which turns long-lived modes into unstable ones \cite{Cardoso:2007az,Chirenti:2008pf,Cardoso:2014sna,Moschidis:2016zjy}. Therefore, constructing stable ECOs which can reproduce astrophysical BH observations seems not to be as simple as finding ultracompact configurations with the potential for counteracting their self-gravity in stationary solutions (which in itself is a highly non-trivial task). In this work we take a novel approach to studying the possible instability of ECOs by means of a non-selfadjoint spectral tool recently introduced~\cite{Jaramillo2020} to the subject area of GW physics: the pseudospectrum of linear operators~\cite{Trefethen:2005,Davie07,Sjostrand2019,Ashida:2020dkc}. 

In particular, ref.~\cite{Jaramillo2020}, followed by \cite{Jaramillo:2021tmt,Destounis:2021lum,Gasperin2021,Jaramillo:2022kuv}, introduced the concept of pseudospectrum in the gravitational context, and first employed it to provide a systematic explanation of the lack of robustness of QNMs against small fluctuations of the underlying BH scattering potential, that were initially reported in refs.~\cite{Aguirregabiria:1996zy,Vishveshwara:1996jgz,Nollert:1996rf,Nollert:1998ys} (see also refs.~\cite{Barausse:2014tra,Konoplya:2022hll,Konoplya:2022pbc} for further examples where the QNM instabilities manifest). Broadly speaking, the pseudospectrum \cite{Trefethen:2005,Davie07,Sjostrand2019} provides a characterization of analyticity properties of an operator's Green function in terms of a kind of topographic map on the complex plane. Contour lines of this topographic map delineate the regions in which eigenvalues can potentially ``migrate" under a system perturbation of size $\epsilon$, the latter regions being denominated $\epsilon$-pseudospectra. With this tool, one is able to reveal whether a non-conservative (more generally, a so-called `non-normal', see e.g. \cite{Trefethen:2005}) system exhibits a spectral instability. Particularly, the QNM spectrum is recovered in the limit $\epsilon\rightarrow 0$, and tightly packed contour lines around these eigenvalues indicate spectral stability. On the other hand, $\epsilon$-contour lines extending far from the eigenvalues indicate spectral instability. In practice, this implies that a minuscule fluctuation in the underlying (QNM) time generator operator can lead to a disproportionate (with respect to the scale of the fluctuation) migration of QNMs in the complex plane, an effect already observed in several seminal works~\cite{Aguirregabiria:1996zy,Vishveshwara:1996jgz,Nollert:1996rf,Nollert:1998ys}. With the help of the pseudospectrum analysis, the Schwarzschild and Reissner-Nordstr\"om BHs have been shown to suffer from such a spectral instability in their overtones~\cite{Jaramillo2020,Gasperin2021,Jaramillo:2021tmt,Destounis:2021lum}, with possible implications also on their fundamental modes when the effective potential is modified at large scales~\cite{Jaramillo2020,Cheung:2021bol}. 

More recently, ref.~\cite{Jaramillo:2022kuv} has further extended the range of applications of the pseudospectrum in the setting of GW physics. Specifically, whereas refs. \cite{Jaramillo2020,Jaramillo:2021tmt,Destounis:2021lum,Gasperin2021}
focus on the application of the pseudospectrum to the analysis of QNM spectral instabilities, ref.~\cite{Jaramillo:2022kuv} rather addresses a set of genuinely dynamical phenomena, namely the possibility of transient growths in linear dynamics driven by non-selfadjoint operators, on the one hand, and the possibility of so-called pseudo-resonances, on the other hand. Indeed, while at late times the linear system dynamics is dominated by the decaying QNMs of the underlying evolution operator, its non-self-adjoint structure may lead to unexpected early-time transient growth that is neither related to non-linear effects nor to resonant-like responses of the system to external forcing (namely, so-called `pseudo-resonances', in addition to actual proper resonant frequencies; see below), as observed in hydrodynamics \cite{Trefethen:1993,Trefethen:2005,Schmi07}. Such transient and pseudo-resonant effects are not captured in the eigenvalues, but are rather encoded in the full pseudospectrum structure and estimated by a set of specific quantities constructed out of it. In the gravitational case, ref.~\cite{Jaramillo:2022kuv} shows the absence of such early-time transient growth for asymptotically flat, spherically symmetric BH spacetimes. This result is independent of the specific underlying potential dictating the wave dynamics, and can be mathematically related to the fact that the field exits through the spacetime boundaries at the horizon and at infinity. Thus, even though the loss of energy through the BH horizon and at the far wave zone is responsible for the non-self-adjoint character of the operator, the same property turns out to preclude early-time transient growth (in spherical symmetry).

Beyond dynamical transients, ref.~\cite{Jaramillo:2022kuv} also emphasizes that the pseudospectrum framework provides an efficient tool to account for the resonant-like dynamical growth of perturbations when the non-self-adjoint linear system is acted upon by external forces. This fact is particularly relevant for the non-linear gravitational interaction, since the first order perturbation solution acts as a source for the second order dynamics~\cite{CunPriMon80}. In particular, pseudo-resonances can be triggered by external forces when the (real) frequencies in the Fourier decomposition of these forces fall close to values where pseudospectrum contour lines of small $\epsilon$ cross the real axis into the unstable half of the complex plane. Thus, the pseudospectrum of the non-selfadjoint second-order perturbation operator (which turns out to have the same form as the linear-order one) can be used to anticipate potential non-linear instabilities, in particular when $\epsilon-$pseudospectrum contour lines for small $\epsilon$ protrude significantly into the part of the complex plane where unstable modes would reside.

Our goal in this work is to analyze the pseudospectrum of the linear operator describing the scalar and gravitational perturbations of a spherically-symmetric ECO with a reflective surface, in search for information regarding spectral instability, transient growth, and the potential for pseudo-resonant phenomena. As an introductory example for these concepts, we first consider the dynamics dictated by the wave equation with a P\"oschl-Teller (PT) potential, together with a reflective boundary condition at some finite radius. This simplified scenario already captures many features we expect to encounter in the behavior of linear perturbations on ECOs. Then, to model actual ECOs we consider a Schwarzschild exterior geometry with a reflective spherical surface above---but arbitrarily close to---the Schwarzschild radius.

The pseudospectra of the aforementioned configurations reveal, first of all, a similar overtone spectral instability to the one present in BH geometries \cite{Jaramillo2020,Jaramillo:2021tmt,Destounis:2021lum}. More importantly, we have observed that the pseudospectrum around the lowest-lying long-lived QNMs of ECOs has an important structural feature: contour lines of equal resonant magnitude protrude into the unstable region of the complex plane. We use this feature of the pseudospectrum to study the possibility of transient growth in the linear homogeneous problem, as well as the magnitude of the pseudo-resonant amplification of sources with (real) frequencies close to the long-lived QNMs, that exist due to stable trapping. Based on the fact that the homogeneous part of the equation governing second-order perturbations can be written in the same form as that of the first-order (therefore their pseudospectra are identical) in an appropriate gauge~\cite{Gleiser:1996yc}, we show that a ``bootstrap" instability can be triggered, which incites a major discrepancy between linear predictions and the full non-linear evolution. Specifically, the linear-order perturbations excite pseudo-resonances at second-order, leading to a possible breakdown of modal analysis and to an eventual non-linear destabilization of the ECO.

This work is structured as follows. Section~\ref{sec:ECOs} introduces the scenarios we will work with in terms of the equations which describe them. This section also introduces the hyperboloidal framework used to formulate the systems as spectral problems of non-self-adjoint operators, as well as the numerical methods employed in their analysis. Then section~\ref{sec:Pseudospectra} reviews the notion of pseudospectrum and the relevant quantities introduced in ref.~\cite{Jaramillo:2022kuv} to assess the early-time transient dynamics. Finally, section~\ref{sec:Results} presents our main results, before our concluding remarks in section \ref{sec:Conclusion}. In what follows we use geometrized units, such that $G=c=1$.

\section{Perturbations of spherically-symmetric ECOs}\label{sec:ECOs}
Perturbation theory in spherically-symmetric backgrounds is typically described by a dynamical field $\Phi$ obeying the wave equation
\begin{equation}\label{perteq}
-\Phi_{,{t}{t}}+\Phi_{,{x}{x}}-V\Phi=0,
\end{equation}
where $t$ is a time coordinate parametrizing the evolution, and $x$ is a radial coordinate defining the spatial domain of integration; $x\in (-\infty, +\infty)$ for the standard tortoise coordinate. As is usual, a comma in the subscript denotes partial differentiation. The potential $V$ depends on the spacetime background and the nature of the perturbation field.

In asymptotically flat BH spacetimes, $V\rightarrow 0$ as $x\rightarrow \pm \infty$, and we solve eq.~\eqref{perteq} with the left- and right-moving boundary conditions
\begin{equation}
\label{BH_BC_time} 
\Phi_{,t}\pm\Phi_{,x}\sim 0,\quad {\rm for} \quad x \rightarrow \pm \infty,
\end{equation}
which impose that the wave be purely ingoing at the horizon and purely outgoing at infinity.

When modeling an ECO, however, this picture changes. As in the case of neutron stars, one needs a complete model of the ECO interior and knowledge of how its matter content interacts with perturbations. Alternatively, one can impose a set of boundary conditions at its surface to effectively model this interaction. For the latter approach, one can simply consider a solution on the domain $x\in [x_o, +\infty)$, where a boundary condition involving a combination of $\Phi(t, x_o)$ and $\Phi_{,x}(t, x_o)$ is imposed at the finite point $x=x_o$ (see e.g. \cite{Maggio:2020jml}). In this work we will make the simple assumption that all incoming waves are completely reflected at the ECO's surface through a Dirichlet boundary condition,
\begin{equation}
\label{ECO_BC_time}
\Phi(t,x_o)=0.
\end{equation}
One may expect that this is a reasonably good approximation to an ECO which resists rapid increases in its mass. Furthermore, boundary conditions which involve partial absorption would require additional assumptions regarding the nature of the ECO and how its size increases with the absorbed energy~\cite{Vellucci2022,Zhong:2022jke}. In the interest of performing a spectral analysis on a static background, such considerations are left outside the scope of the present work.

The spectral problem is studied in the frequency domain, where the Fourier decomposition $\Phi(t,x)\sim \phi(x) e^{i\omega t}$ yields the eigenvalue equation
\begin{equation}\label{pertF}
{\phi}_{,{x}{x}}+(\omega^2-V)\phi=0.
\end{equation}
QNMs are solutions of eq.~\eqref{pertF} which satisfy purely outgoing boundary conditions at infinity and appropriate boundary conditions at the ECO's surface. Particularly, the Fourier space equivalent of the BC \eqref{BH_BC_time} at infinity, and of \eqref{ECO_BC_time} at the surface, which read
\begin{eqnarray}
\label{BC_Freq}
\phi(x) \sim e^{- i\omega x} \quad &{\rm for}& \quad x\rightarrow \infty, \\
  \phi(x_o) = 0.  \quad &{\rm at}& \quad x=x_o.
\end{eqnarray}

\subsection{Hyperboloidal foliation}
In this work, we follow refs.~\cite{Zenginoglu:2007jw,Zenginoglu:2011jz,Zenginoglu:2021tvv,Ansorg2016,PanossoMacedo:2018hab,PanossoMacedo:2020biw,Jaramillo2020} in incorporating the asymptotic boundary conditions via the hyperboloidal approach to perturbation theory. By exploiting the freedom in the choice of coordinates, we deform the time slices and compactify the radial direction via
\begin{equation}
\label{Hyp_Transf}
t = \lambda \left(\tau - H(\chi) \right),\quad x = \lambda \, g(\chi),
\end{equation}
with $H(\chi)$ the so-called height function, which ensures that the wave zone (future null infinity) is approached as $x\rightarrow \infty$, $g(\chi)$ a function mapping $x\in [x_o,\infty)$ into the compact interval $\chi\in[\chi_{_{\scri^+}}, \chi_o]$, and $\lambda$ a characteristic (constant) length scale of the problem.

The wave equation \eqref{perteq} takes the form \cite{Jaramillo2020,Destounis:2021lum}
\begin{equation}\label{hyperboloidal}
	-\Phi_{,\tau\tau}+L_1\Phi+L_2\Phi_{,\tau}=0,
\end{equation}
where
\begin{equation}
\label{eq:Operators_L1_L2}
	\begin{split}
		L_1&=\frac{1}{w(\chi)}\bigg[\partial_\chi \left(p(\chi) \partial_\chi \right) - \hat{V}(\chi) \bigg],\\
		L_2&=\frac{1}{w(\chi)}\bigg[2\gamma(\chi)\partial_\chi + \partial_\chi\gamma(\chi)\bigg],
	\end{split}
\end{equation}
and the corresponding functions
\begin{eqnarray}
\label{eq:funcs_wave}
p = \dfrac{1}{|g'|}, \, \gamma(\chi) = p h', \, w = \dfrac{1}{p} \left( 1 - \gamma^2\right), \, \hat{V} = \dfrac{\lambda^2}{p} V, 
\end{eqnarray}
where  the prime denotes the derivative with respect to $\chi$.

Equation \eqref{hyperboloidal} can be written as a matrix evolution problem through a first-order reduction in time by introducing $\psi=\phi_{,\tau}$ such that
\begin{equation}\label{matrix evolution}
	\dot u=i L u, \quad 	L =\frac{1}{i}\!
	\left(
	\begin{array}{c  c}
		0 & 1 \\
		L_1 & L_2
	\end{array}
	\right), \quad u=\left(
	\begin{array}{c}\phi \\ \psi \end{array}\right).
\end{equation}
Given initial data $u(0,\chi)$, eq.~\eqref{matrix evolution} has the formal solution
\begin{equation}
	\label{evolution_operator}
	u(\tau,\chi)=e^{iL\tau}u(0,\chi),
\end{equation}
in terms of the \emph{evolution operator} $e^{iL\tau}$. By further performing a harmonic decomposition $u(\tau,\chi)\sim \hat{u}(\chi)e^{i\omega\tau}$ in eq.~\eqref{matrix evolution} we arrive at the eigenvalue equation
\begin{equation}\label{eigenvalue problem}
	L \hat u_n=\omega_n \hat u_n,
\end{equation}
where, imposing the  surface boundary condition \eqref{ECO_BC_time}, $\omega_n$ becomes an infinite set of QNMs of the operator $L$, with $n\geq 0$ the overtone number.

Note that the outgoing boundary condition \eqref{BH_BC_time} for $x\to\infty$ is automatically imposed by the hyperboloidal slicing. Indeed, the main motivation for employing the hyperboloidal approach is the geometric imposition of outgoing boundary conditions at future null infinity (and at the event horizon of BH spacetimes). With the light cones pointing outwards from the integration domain, the causal degrees of freedom must satisfy the regularity conditions underlying the wave equation. In practical terms, this property manifests via $p(\chi_{_{\scri^+}})=0$, i.e. the term proportional to the second spatial derivative vanishes, and the remaining term imposes the appropriate condition for the relation between $\phi(\chi_{_{\scri^+}})$ and $\phi_{,\chi}(\chi_{_{\scri^+}})$.

At the ECO boundary $\chi_o$, however, the wave equation does not contain any singularity, and we are free to impose any desired external boundary condition. In particular, for our simple ECO model we choose to impose eq.~\eqref{ECO_BC_time}, or equivalently eq.~\eqref{BC_Freq}, which translates into
\begin{equation}\label{BC_FirstOrderRed}
u(\chi_o, \tau) = \hat u(\chi_o) = 0.
\end{equation}

As in ref.~\cite{Jaramillo2020}, we explicitly construct working examples for the cases of dynamics dictated by either the PT potential or the external Schwarzschild spacetime potential. While the former provides us with a simple toy model to exemplify the ideas presented in the work and set a benchmark for the results, the latter has direct astrophysical relevance when modeling ECOs.

\subsubsection{P\"oschl-Teller Potential}
Let us first analyze the simpler case of eq. \eqref{perteq} with the PT potential,
\begin{equation}
V(x)=\frac{V_0}{\cosh^2(x/\lambda)},
\end{equation}
where $V_0$ is a positive constant. We will use units $V_0=1$ by setting the length scale to $\lambda=\sqrt{V_0}$ in transformations \eqref{Hyp_Transf}. The hyperboloidal coordinates adequate for the asymptotic behavior of this potential are \cite{Jaramillo2020}
\begin{equation}
\label{PT_hyp_coord}
t=\lambda \bigg[ \tau-\frac{1}{2}\ln(1 - y^2)\bigg],\quad {x}= \lambda \, \text{arctanh}(y),
\end{equation}
where $y\in[-1,1]$. Rather than using the whole range of $y$ for our toy ECO construction, we follow ref.~\cite{Price:2017cjr} and place a Dirichlet boundary condition at $y_o= 1-\mathcal{E}$. 

It is, thus, more convenient to introduce a new spatial coordinate $\chi\in[-1,1]$ via
\begin{equation}
\label{PT-ECO_coord}
y =   y_o\dfrac{1+\chi}{2} - \dfrac{1-\chi}{2}=\chi-\frac{\mathcal{E}}{2}(1+\chi),
\end{equation}
which maps the PT-ECO boundary $y_o$ to $\chi_o = 1$, while keeping future null infinity $y_{_{\scri^+}}= -1$ at $\chi_{_{\scri^+}} = -1$. Inserting \eqref{PT-ECO_coord} into \eqref{PT_hyp_coord}, and making use of the definitions in eqs.~\eqref{eq:funcs_wave} one obtains the functions
\begin{eqnarray}
\label{eq:p_PT}
p(\chi)&=& 1-\chi^2 +\dfrac{\mathcal{E}}{2} (1+\chi)^2, \\
\label{eq:gamma_PT}
\gamma(\chi) &=& - \chi + \dfrac{\mathcal{E}}{2}(1+\chi), \\
\label{eq:w_PT}
 w(\chi) &=& \hat V(\chi) = 1 - \dfrac{\mathcal{E}}{2},
\end{eqnarray}
from which the operators $L_1$ and $L_2$ follow directly. 

As already mentioned, $p(\chi)$ vanishes at $\chi_{_{\scri^+}} = -1$, but not at $\chi_o=1$ when $\mathcal{E}\neq 0$. Thus, in this case we have the freedom to specify the boundary condition at the PT-ECO surface. Furthermore, we recover the expressions from ref.~\cite{Jaramillo2020} in the limit ${\cal E}\rightarrow 0$, in which $\chi=1$ also corresponds to an asymptotic region.

\subsubsection{Regge-Wheeler Potential}
A similar set of hyperboloidal coordinates is introduced for perturbations propagating on spherically-symmetric ECO of mass $M$ with a surface located at a radial position 
\begin{equation}
\label{eq:ECO_surface}
r_0=(1+\mathcal{E})r_{\rm h}, \quad \mathcal{E}\ll 1,
\end{equation}
where $r_h=2M$. The geometry outside the ECO is described by the Schwarzschild metric. In the standard Schwarzschild coordinates $(t,r,\theta, \varphi)$, the line element reads
\begin{equation}\label{Schw}
{\rm d}s^2=-f(r){\rm d}t^2+f(r)^{-1}{\rm d}r^2+r^2{\rm d}\Omega^2_2,
\end{equation}
where $f(r)=1-r_{\rm h}/r$ and ${\rm d}\Omega^2_2$ is the line element of the unit $2$-sphere. The radial coordinate $x$ defining the spatial domain in eq.~\eqref{perteq} is the tortoise coordinate
\begin{equation}
\label{eq:tortoise_coord}
x =r+r_{\rm h}\log \left(\frac{r}{r_{\rm h}}-1\right),
\end{equation}
whereas the potential $V$ follows from the Regge-Wheeler-Zerilli formalism \cite{Berti:2009kk}. In this work, as an example, we focus on the Regge-Wheeler potential
\begin{equation}\label{potential}
V=\dfrac{f(r)}{r^2}\left(\ell(\ell+1)+(1-s^2)\frac{r_{\rm h}}{r}\right),
\end{equation}
with $\ell$ the angular number and $s$ the spin parameter characterizing the scalar ($s=0$), vector ($\pm1$) and tensorial ($\pm2$) sectors of the metric perturbation.

We begin with the hyperboloidal coordinates $(\tau,\sigma, \theta, \varphi)$ introduced in \cite{Ansorg2016,Jaramillo2020} 
\begin{equation}
\label{eq:HyperbCoord_Sch}
t = \lambda\Bigg[ \tau - \dfrac{r_{\rm h}}{\lambda}\left( \ln(\sigma) + \ln(1-\sigma) - \sigma^{-1} \right)  \Bigg], \, r = \dfrac{r_{\rm h}}{\sigma},
\end{equation}
which map the BH horizon $r=r_{\rm h}$ to $\sigma =1$ and future null infinity to $\sigma =0$. Then, according to eq.~\eqref{eq:ECO_surface}, the ECO surface $r_o$ is located at 
\begin{equation}
\sigma_o = \dfrac{r_{\rm h}}{r_o} = \bigg( 1 + \mathcal{E} \bigg)^{-1} \approx 1 - \mathcal{E}, \quad {\cal E}\ll 1.
\end{equation}

Similar to the PT case, we introduce a final coordinate $\chi\in[0,1]$ via 
\begin{equation}
\label{eq:ECO_CompactCoord}
\sigma = \sigma_o \chi.
\end{equation}
Thus, we retain the location of future null infinity at $\chi_{_{\scri^+}} = 0$, while mapping the ECO surface to $\chi_o = 1$.

The transformations \eqref{eq:HyperbCoord_Sch} and \eqref{eq:ECO_CompactCoord}, combined with eq.~\eqref{eq:tortoise_coord} yield the functions
\begin{eqnarray}
\label{eq:p_Schwarzschild}
p(\chi) &=& \dfrac{\lambda}{r_{\rm h} } \dfrac{\chi^2 \left(1+ {\cal E} - \chi \right)}{(1+ {\cal E} )^2}, \\
\label{eq:gamma_Schwarzschild}
\gamma(\chi)  &=& 1- \dfrac{2 \chi^2}{(1+  {\cal E} )^2}, \\
w(\chi)  &=& \dfrac{4 r_{\rm h}}{\lambda}\dfrac{(1+{\cal E} +\chi)}{(1+{\cal E})^2}, \\
\label{eq:V_Schwarzschild}
\hat V(\chi) &=& \dfrac{\lambda}{r_{\rm h}} \left[ \frac{\ell(\ell+1)}{1+{\cal E}} + \frac{(1-s^2)\chi}{(1+{\cal E})^2} \right].
\end{eqnarray}

Once again we emphasize the singular character of the equation at future null infinity, captured by the property $p(0) = 0$. Imposing the regularity of the perturbation field there, automatically fixes the necessary boundary condition. At $\chi=1$, the equation does not degenerate, and we are free to impose the boundary conditions fixed by eq.~\eqref{BC_FirstOrderRed}. This freedom is no longer available for ${\cal E}\rightarrow 0$. In this limit, eqs.~\eqref{eq:p_Schwarzschild}-\eqref{eq:V_Schwarzschild} reduces to the corresponding expression for the Schwarzschild BH spacetime~\cite{Jaramillo2020}, in which $p(1)$ also vanishes.

This compactified hyperboloidal approach allows us to cast the relevant equations in terms of the spectral problem of a non-selfadjoint operator, and to employ the corresponding tools that recover the results from the theory of scattering resonances (see refs. in \cite{Jaramillo2020,Jaramillo:2021tmt}). Section \ref{sec:Pseudospectra} introduces several pseudospectrum (or `non-modal analysis'~\cite{Trefethen:1993,Trefethen:2005,Schmid:2007}) quantities which we will use in this context. Hereafter, we solve the underlying equations and evaluate these quantities numerically. The following section presents the numerical methods employed.

\subsection{Numerical Method}\label{sec:NumMethod}
We employ a collocation spectral method with the Chebyshev polynomials as a basis~\cite{Trefethen2000,boyd2001chebyshev,Trefethen:2005,canuto2007spectral}. To that end, we fix a numerical resolution $N$ (number of spatial discretization points) and introduce the Chebyshev-Lobatto grid
\begin{equation}
\label{eq:Lobatto-Grid}
\chi_j = \cos(j\, \pi/N), \quad j = 0 \cdots N,
\end{equation}
which allows the discretization of the differential operators from eqs.~\eqref{eq:Operators_L1_L2} in terms of their corresponding spectral approximation~\cite{Trefethen2000,boyd2001chebyshev,canuto2007spectral,Trefethen:2005} (when the interval $\left[\chi_{_{\scri^+}},\chi_o\right]$ is different form $[-1,1]$, we employ a linear rescaling). The resulting matrix $L$ in eq.~\eqref{matrix evolution} has dimensions $\mathcal{N}=2(N+1)\times 2(N+1)$.

The reflective boundary condition at the ECO surface is subsequently imposed by removing the row and column of the discretized matrix $L$ corresponding to the position of the surface in the discrete spatial grid (see e.g. \cite{Trefethen2000}). Then it follows straightforwardly from \eqref{eigenvalue problem} that the spectrum $\sigma(L)$ of this discretized and dimensionally-reduced (due to the reflective boundary condition) $L$ consists of the QNMs of the ECO with purely outgoing boundary conditions at future null infinity and a Dirichlet boundary condition at its surface.

We note that we explored several different methods of imposing the necessary boundary conditions, before finding that the above is the most efficient for numerical computation. For instance, to include future null infinity into the computational domain, and thus impose the outgoing condition there through regularity, a spacetime slicing can also be performed along the null outgoing Eddington-Finkelstein coordinate $u=t-x$ (toward which the hyperboloidal $\tau$ tends at infinite radius). In BH spacetimes, this slicing method is not viable because the limit $r\rightarrow r_h$ along the surface $u=$const. leads to the white-hole horizon, where the ingoing boundary condition for QNMs is not compatible. However, one encounters no problem in the ECO setup because the domain is cut at the ECO surface $x_o$. On the other hand, to impose boundary at ECO surface, we have explored the alternative methods described in refs.~\cite{Trefethen2000,Jackiewicz:2002,Trefethen:2005} as well as the possibility of enforcing the boundary condition at the ECO surface via an auxiliary field $u = (1-\chi) w$.

\section{Pseudospectrum and non-modal analysis: dynamical transients}\label{sec:Pseudospectra}

In this section we introduce the mathematical concepts relevant to our study of ECOs, closely following the discussion in \cite{Jaramillo:2022kuv}. In particular, we provide a brief introduction to the notion of pseudospectrum (for more detail see e.g. refs.~\cite{Trefethen:2005,Jaramillo2020} and references therein) and we review the concepts of numerical abscissa and Kreiss constant. As we will explain, these quantities allows us to obtain information regarding non-modal transient effects, as introduced to the context of GW physics in~\cite{Jaramillo:2022kuv}.

\subsection{Pseudospectrum}
Before we introduce the pseudospectrum of an operator $L$, let us first recall that the operator's spectrum $\sigma(L)$ is understood as the set of complex values $\lambda$ where the so-called resolvent
\begin{equation}\label{pseudoDLg}
R_L(\lambda) = (L - \lambda \mathbb{I})^{-1},
\end{equation}
(with $\mathbb{I}$ the identity operator) is not defined as a bounded operator. 
Essentially, the resolvent is given by the Green function of the eigenvalue problem $(L -\lambda\mathbb{I}) u=0$ . This definition generalises our practical experience with finite matrices, where the spectrum is obtained by finding discrete $\lambda$-values such that 
$\det\left[L - \lambda \mathbb{I}\right] = \det \left[ R_L(\lambda)^{-1} \right] =0$.

Building upon this notion, the $\epsilon$-pseudospectrum $\sigma^\epsilon(L)$ is characterised by the set of complex values $\lambda$ such that $\|R_L(\lambda) ^{-1}\| < \epsilon$. Two aspects are worth mentioning: (i) the spectrum $\sigma(L)$ is recovered when $\epsilon \rightarrow 0$, and (ii) in contrast to the spectrum, the $\epsilon$-pseudospectrum depends on the choice of a scalar product~\cite{Gasperin2021}. We will discuss the latter in more detail in sec.~\ref{sec:energynorm}.

In fact, it can be shown that the following definitions of the pseudospectrum are equivalent~\cite{Trefethen:2005}
\begin{eqnarray}\label{pseudoDL}
\label{eq:pseudospectra_def1}
\sigma^\epsilon(L)&=&\{\lambda\in\mathbb{C}:\|R_L(\lambda)\|>1/\epsilon\}, \\
                             &=&\{\lambda\in\mathbb{C},\exists\delta L,\|\delta L\|<\epsilon:\lambda\in\sigma(L+\delta L)\}, \\
\label{eq:pseudospectra_def3}
			   &=& \{\lambda\in\mathbb{C},\exists v\in\mathbb{C}^n:\|Lv-\lambda v\|<\epsilon\|v\|\}.
\end{eqnarray}
The first is the aforementioned one, where the norm of the resolvent provides an $\epsilon-$contour level bounding the set $\sigma^\epsilon(L)$. This prescription is particularly useful for numerical computations. 

The second definition illustrates how the spectral instability under a perturbation $\delta L$ is encoded in the pseudospectrum. In other words, when a perturbation of magnitude $\epsilon$ is added to the operator $L$, the perturbed spectrum $\sigma(L+\delta L)$ can be displaced only within the confines of the $\epsilon$-pseudospectral contour lines, and the $\epsilon$-pseudospectrum can in fact be defined as the set of all possible displacements of the spectrum under all possible perturbations of magnitude equal or smaller than $\epsilon$.

Finally, the third definition introduces the concept of a pseudo-eigenmode $v$ of the operator, which is connected to the notion of \emph{pseudo-resonance}: a point in the complex plane (not coincident with the spectral points) where the resolvent norm has a very large value.

Upon calculating the pseudospectrum, a general criterion for visually interpreting the result is the following: for each value of $\epsilon>0$, if $\sigma^\epsilon(L)$ is a set of circular regions (``tubular sets'') of radius $\epsilon$ around $\sigma(L)$ in the complex plane, then the operator can be said to be spectrally stable (see e.g., fig. 4 in \cite{Jaramillo2020} and fig. 10 in \cite{Destounis:2021lum}). Otherwise, if $\sigma^\epsilon(L)$ encompasses regions in the complex plane much larger than these ones, then $L$ is spectrally unstable, as has been found for the QNM operators in the Schwarzschild and Reissner-Nordstr\"om BHs \cite{Jaramillo2020,Destounis:2021lum,Jaramillo:2021tmt}.

As discussed in the introduction and in further detail in refs.~\cite{Trefethen:1993,Trefethen:2005,Schmid:2007,Jaramillo:2022kuv}, the  pseudospectrum contains more information than just the characterization of the spectral stability of $L$. In the $\epsilon\to 0$ limit, it determines the late-time evolution of the problem through the QNM spectrum,\footnote{We restrict the discussion to the discrete part of the spectrum $\sigma(L)$ yielding the QNM ring-down behaviour. As it is well known for asymptotically flat BH spacetimes, one also encounters a very late time power-law decay \cite{Gundlach:1993tp}. In the current context, this tail is understood in terms of the continuous part of the spectrum $\sigma(L)$ \cite{Leaver86c,Ansorg2016,PanossoMacedo:2018hab}} but away from this limit it provides information regarding the early and intermediate-time evolution of the perturbations, as well as the existence and magnitude of possible pseudo-resonances, which the eigenvalue analysis fails to identify.

\subsection{Transient growth of linear perturbations}\label{sec:Transient}
Following closely the ideas put forward in \cite{Jaramillo:2022kuv} in the gravitational setting (and more generally in refs.~\cite{Trefethen:1993,Trefethen:2005,Schmid:2007}), we will briefly present the quantities related to evolution operator $e^{i L \tau}$ which allow us to obtain information regarding the early-time evolution of perturbations. We recall that the dynamics of the system is given by the wave equation expressed in the form \eqref{matrix evolution}, and the evolution operator arises when expressing the formal solution to the system~\eqref{evolution_operator} in terms of initial data $u(0, \chi)$. In particular, the operator's norm $\|e^{i L \tau} \|$ monitors the maximum growth rate of solutions $u(\tau)$ in time.\footnote{For simplicity, in this section we omit the solution's dependence on the spatial coordinate $\chi$.}

In the standard modal analysis of the dynamical stability of the wave equation, one typically focuses on the fundamental QNM $\omega_0$ --- particularly, on the sign of its imaginary part. Modal stability is shown when ${\rm Im}(\omega_0)$ takes on values in the `stable' half of the complex plane, i.e. ${\rm Im}(\omega_0)>0$ with the convention used in this work. 
The quantity typically used to monitor this late-time stability is the so-called \emph{spectral abscissa}~\footnote{Note the $-\pi/2$ rotation as compared with the standard `abscissa', due to the `$i$' factor in the definition of the infinitesimal time generator, as compared with the notation in e.g. \cite{Trefethen:2005}, namely $\alpha(iL) = \sup_{z\in\sigma(iL)} {\rm Re}(z)$. This accounts for the minus sign in the definition (see~\cite{Jaramillo:2022kuv} for a more complete discussion).}
\begin{equation}
\label{e:spectral_abscissa}
\alpha(L) = -\inf_{z\in\sigma(L)} {\rm Im}(z),
\end{equation}
which formally identifies ${\rm Im}(\omega_0)$ (when a spectral gap exists between $\omega_0$ and the real line).

For self-adjoint operators $\|e^{i L \tau} \|\sim e^{ \alpha(L)\tau}$, implying that the growth rate of solutions $u(\tau)$ is  well estimated by the decaying exponential $e^{- \tau {\rm Im} (\omega_0)}$. The picture changes when $L$ is not self-adjoint. In this case, the behavior $\|e^{i L \tau} \|\sim e^{- \tau {\rm Im} (\omega_0)}$ manifests only at late times. Throughout the rest of the evolution, the spectral abscissa only provides the lower bound
\begin{equation}
\label{eq:spec_abscissa_lowbound}
\|e^{iL\tau}\| \geq e^{\alpha(L)\tau} = e^{- \tau {\rm Im} (\omega_0)},\quad \forall \tau\geq0,
\end{equation}
leaving open the possibility of initial transient growth, even if at late times the solutions still decays according to ${\rm Im}(\omega_0)>0$. 

The possibility for transient growth has far-reaching implications in physics, where the linear evolution is often only an approximation, the validity of which relies on the absence of such growth. For instance, in hydrodynamics \cite{Trefethen:1993} the accumulation of transient growth in perturbations of certain configurations has been shown to lead to a breakdown of the linear approximation, from where non-linear instabilities take over --- all this in a system where an eigenvalue analysis would predict stability.

It is therefore important to have a full grasp of the evolution of system described by \eqref{evolution_operator} at different times. Specifically, one can identify three distinct phases to the evolution dictated by an operator with stable eigenmodes (${\rm Im}(\omega_0)>0$): a possible (i) initial growth, reaching a (ii) transient maximum, before proceeding to the (iii) late-time decay. Specific tools are available to monitor each of these stages. While stage (iii) is described by the spectral abscissa $\alpha(L) = {\rm Im}(\omega_0)$, we will now introduce the quantities which provide information on stages (i) and (ii).

\subsubsection{Numerical abscissa}
The \emph{numerical abscissa} $\omega(L)$, that is defined in terms of the so-called `numerical range' of $L$ (see \cite{Trefethen:2005,Jaramillo:2022kuv}), characterizes the initial slope of the dependence in time of the norm of the evolution operator
\begin{equation}
  \label{e:init_growth}
\omega(L)=\left.\frac{\text{d}}{\text{d}\tau}\|e^{iL\tau}\|\right|_{\tau=0}.
\end{equation}
Therefore, a value $\omega(L)>0$ indicates the presence of an initial growth in the evolution. This quantity can be related to the $\epsilon-$pseudospectrum in the limit $\epsilon\rightarrow \infty$~\cite{Jaramillo:2022kuv,Trefethen:2005}. Moreover, this quantity also provides an upper bound on the evolution operator's norm at any time $\tau>0$ through the relation
\begin{equation}\label{nabound}
\|e^{iL\tau}\|\le e^{\omega(L)\tau},\quad \forall \tau\geq0.
\end{equation}
Particularly, one sees that when $\omega(L)=0$ the solution of the evolution problem cannot grow and, generically, its norm decreases in time (the operator $e^{iL\tau}$ in this case is said to be \textit{contractive}, namely $\|e^{iL\tau}\|\leq 1, \forall \tau>0$), whereas when $\omega(L)>0$ the possibility of transient growth remains.

For practical purposes, the numerical abscissa can also be characterized and computed \cite{Jaramillo:2022kuv} through the eigenvalue the problem
\begin{equation}\label{eq:numericalabsicissa_eignvalue}
\dfrac{1}{2i}\left(L^\dagger - L\right) u_{L} = \lambda u_{L}.
\end{equation}
In particular, the eigenvalues $\lambda$ of this problem are real numbers (by construction, since the operator in the left-hand side is selfadjoint), and $\omega(L)$ is simply the largest one.

As mentioned above, the spectral and numerical abscissa are related to the $\epsilon\rightarrow 0$ and $\epsilon\rightarrow \infty$ limits of the $\epsilon-$pseudospectra, respectively. Given the dynamical stages of the evolution they are each related to, a time scale $\tau_{\epsilon}\sim 1/\epsilon$ was introduced in ref.~\cite{Jaramillo:2022kuv}  to provide an estimation of the relation between different moments of the evolution governed by the operator \eqref{evolution_operator} and the $\epsilon$-pseudospectral contours.

\subsubsection{Pseudospectral abscissa and Kreiss constant}
 
To estimate the maximum of the possible dynamical transient, that is controlled by the maximum growth of the norm of the evolution operator at intermediate times,
we can utilize the so-called \emph{pseudospectral abscissa} of $L$, defined in our setting as~\footnote{Again, as in eq. \eqref{e:spectral_abscissa}, the minus sign follows from the factor ``$i$'' in the infinitesimal time generator.} 
\begin{equation}
\label{eq:pseudospectra_abscissa}
\alpha_{\epsilon}(L) = -\inf_{z\in\sigma^\epsilon(L)} {\rm Im}(z).
\end{equation}
This quantity allows us to put a sharper lower bound on the evolution norm than eq.~\eqref{eq:spec_abscissa_lowbound}, giving a lower bound to a possible transient peak. This bound is obtained from the $\epsilon$-pseudospectrum at intermediate values of $\epsilon$ through the relation
\begin{equation}\label{bound}
\sup_{\tau\ge 0}\|e^{iL\tau}\|\ge\frac{\alpha_\epsilon(L)}{\epsilon},\quad \forall\epsilon>0.
\end{equation}
In other words, if the $\epsilon-$pseudospectral contour lines protrude {\em significantly} (as quantified by $\alpha_\epsilon(L)>\epsilon$) into the `unstable' half of the complex plane, then transient growth occurs in the evolution and the peak corresponds to the intermediate value of $\epsilon$ that maximizes the right-hand side of eq. \eqref{bound}. This naturally leads, upon  maximizing eq.~\eqref{bound} over $\epsilon >0$, to the absolute lower bound
\begin{equation}
\label{eq:kreiss_bound}
\sup_{t\ge 0}\|e^{iL\tau}\|\ge {\cal K}(L)
\end{equation}
in terms of the so-called \emph{Kreiss constant}
\begin{eqnarray}\label{kreiss}
\mathcal{K}(L)&=&\sup_{\epsilon>0}\frac{\alpha_\epsilon(L)}{\epsilon} \\
&=&\sup_{{\rm Im}(z)<0} \left\{ |{\rm Im}(z)| \,\cdot \| R_{L}(z) \|\right\}, \nn
\end{eqnarray}
where the characterization in the second line (cf. \cite{Trefethen:2005,Jaramillo:2022kuv}) provides an efficient approach to its calculation. If the fundamental QNM were unstable, i.e. if $\mathrm{Im}(\omega_0)<0$ and therefore $\alpha(L)=\lim_{\epsilon\to 0}\alpha_\epsilon(L)>0$, then the Kreiss constant $\mathcal{K}(L)$ would be directly divergent (as follows from the first line in \eqref{kreiss} by taking $\epsilon\to 0$),  consistently with the unbounded growth such a mode would produce. On the other hand, if $\alpha(L)\le 0$, then ${\cal K}(L)$ can be either larger than 1, in which case there would be necessarily a transient growth, or equal to 1. In the latter case, no conclusion can be drawn in general, since \eqref{eq:kreiss_bound} saturates and trivializes for $\tau=0$ and the inequality is not significant. However, if the numerical abscissa actually vanishes $\omega(L)=0$ (indicating through \eqref{e:init_growth} the absence of an initial growth), then the Kreiss constant must indeed be unity ${\cal K}(L)=1$
(cf. discussion \cite{Jaramillo:2022kuv}) and in this (contractive) case we know a priori
that no transient happens along the evolution.

At a practical level, the calculation of the ratio $\alpha_\epsilon(L)/\epsilon$ comes ``for free" once the pseudospectrum is known. All one needs to do is take the point of an $\epsilon$-pseudospectrum contour line with minimum imaginary part. Particularly, lines which protrude into the negative imaginary half-plane are the ones which will provide a sensible lower bound. The supremum corresponding to the Kreiss constant is then reached somewhere between the first such protruding $\epsilon$-contour and the outermost one, i.e. $\epsilon\to\infty$.

Note that all the quantities discussed in this section rely on specifying a norm for the operators involved. When calculating the pseudospectrum, and quantities derived therefrom, this norm is used to evaluate $\| R_{L}(z) \|$. For the explicit calculation of the numerical abscissa via eq.~\eqref{eq:numericalabsicissa_eignvalue}, that assumes a norm associated with a scalar product, such scalar product enters in the very construction of the adjoint operator $L^{\dagger}$. We now proceed to define the norm, and the scalar product it is associated with, which we will employ in this work.

\subsection{Energy norm}\label{sec:energynorm}

To choose the norm with which to evaluate the above quantities, we follow past works~\cite{Jaramillo2020,Jaramillo:2021tmt,Destounis:2021lum,Gasperin2021} in appealing to a physical argument for the measurement of the magnitude of the perturbations described by \eqref{perteq}.
Particularly, we will employ the \emph{energy norm} (see also \cite{Driscoll:1996}), which quantifies this magnitude through its contribution to the system's energy (for an in-depth discussion of the role of the norm and, in particular,
the energy norm, see \cite{Gasperin2021}).

The energy associated to the state vector $u=\left(\phi, \psi \right)^T$ satisfying the wave equation \eqref{matrix evolution} reads
\begin{equation}\label{energy norm}
\begin{split}
&E(\phi, \psi) = \| u \|^2_{_E}  \\
 &=\frac{1}{2}\int_{\chi_{_{\scri^+}}}^{\chi_o} \left(\omega(\chi)|\psi|^2
+ p(\chi)|\partial_\chi\phi|^2 + \hat{V}(\chi) |\phi|^2\right) d\chi.
\end{split}
\end{equation}
The subscript $E$ denotes the energy norm and the integration limits correspond to the compactified spatial interval between future null infinity and the ECO surface in our hyperboloidal scheme. An energy scalar product actually can be introduced as the quadratic form associated to such energy norm, in the form
\begin{eqnarray}
\label{eq:EnergyScalarProduct}
&&	\langle u_1,\! u_2\rangle_{_{E}} = \Big\langle\begin{pmatrix}
	\phi_1 \\
	\psi_1
\end{pmatrix}, \begin{pmatrix}
	\phi_2 \\
	\psi_2
\end{pmatrix}\Big\rangle_{_{E}}  \\
&&=
\frac{1}{2} \int_{\chi_{_{\scri^+}}}^{\chi_o} \left( \omega(\chi)\bar{\psi}_1 \psi_2 + p(\chi)  \partial_\chi\bar{\phi}_1\partial_\chi\phi_2 + q(\chi)\bar{\phi}_1 \phi_2 \right)  d\chi, \nn\label{energy scalar product}
\end{eqnarray}
such that, by construction, $\|u\|^2_{_{E}} = \langle u, u\rangle_{_{E}}$. With eq.~\eqref{eq:EnergyScalarProduct} we can evaluate the pseudospectrum according to definition \eqref{eq:pseudospectra_def1}, and we can also analytically deduce the expression for the adjoint $L^\dagger$~\cite{Jaramillo2020,Gasperin2021}.

More specifically, the $\epsilon$-pseudospectrum in the energy norm $\sigma^\epsilon_{_E}(L)$ can be calculated as the set
\begin{equation}
\label{pseudospectra energy norm}
\sigma^\epsilon_{_E} (L) = \{\omega\in\mathbb{C}: s_{_E}^\mathrm{min}(\omega \mathbb{I} - L)<\epsilon\}, 
\end{equation}
with $s_{_E}^\mathrm{min}$ the minimum of the generalized singular value decomposition
\begin{equation}
\label{svd definition}
s_{_E}^\mathrm{min}(\mathcal{M}) = \min \{\sqrt{\omega}:  \omega\in \sigma(\mathcal{M}^\dagger \mathcal{M}) \}.
\end{equation}
Appendix C.3 of ref. \cite{Jaramillo2020} details the calculation of the discretized energy scalar product within the numerical scheme presented in sec.~\ref{sec:NumMethod}. 

By following \cite{Jaramillo2020,Gasperin2021}, we also derive an analytic expression for the formal adjoint $L^\dagger$
\begin{equation}
\begin{split}
\expval{L^\dagger u_1,u_2} &= \expval{u_1,Lu_2}   \\
&= \expval{L u_1,u_2}  + \expval{ L^\partial u_1,u_2},
\end{split}
\end{equation}
with $L^\partial$ yielding a boundary contribution in the form
\begin{equation}
\expval{ L^\partial u_1,u_2} = i \left[ 2  \gamma\bar\psi_1 \psi_2 + p \left(\bar\phi_1' \psi_2 + \phi_2' \bar\psi_1 \right) \right] \bigg |_{\chi_{_\scri^+}}^{\chi_o}.
\end{equation}
We emphasize that the term $\expval{ L^\partial u_1,u_2}$ is precisely the one responsible for the non-self-adjoint character of the operator $L$.

At future null infinity, the only contribution to $L^\partial$ comes from the term which contains $\gamma(\chi)$, since $p(\chi)$ always vanishes at $\chi_{_{\scri^+}}$. At the ECO surface, $\expval{ L^\partial u_1,u_2}$ vanishes altogether as a consequence of the boundary condition~\eqref{BC_FirstOrderRed}. This result makes apparent the fact that the reflective ECO surface does not contribute to the energy outflow which makes the operator non-self-adjoint; the only such contribution comes form waves propagating away towards infinity, and is the same as in the BH case \cite{Jaramillo2020}. We can thus express the boundary operator $L^\partial$ in terms of a Dirac-delta distribution 
\begin{equation}
\label{eq:BoundaryOp}
L^\partial={i}\begin{pmatrix}
0 & 0\\ 0 & -2\gamma  \delta(\chi - \chi_{_{\scri^+}})
\end{pmatrix}.
\end{equation}
Following previous works~\cite{Jaramillo2020,Jaramillo:2021tmt,Destounis:2021lum,Gasperin2021}, we have performed consistency tests that confirm the qualitative robustness of the results presented in the following sections against modification of the norm in use. Particularly, we have examined several options which retain the form \eqref{energy norm} with different multiplication weights in the integral, with one of these options being the standard $L^2$ norm of the order-reduced system. We find no qualitative difference in the resulting pseudospectrum. An open question (beyond the scope of this work) is to study the pseudospectra of norms specially constructed to address specific mathematical issues in the QNM problem~\cite{Gasperin2021,Gajic:2019qdd,Gajic:2019oem,Sjostrand2019,galkowski2021outgoing}, as well as aspects related to the ``energetic content'' of the norm~\cite{Gasperin2021,Zimmerman:2014aha,Yang:2022wlm} used to construct the pseudospectrum.

\section{Results}\label{sec:Results}
We now apply the elements of the pseudospectrum framework introduced in the previous section to the equations governing the dynamics of perturbations on ECO spacetimes. As an example and motivation, we first study the toy model of a wave equation with the (modified) PT potential. This model already captures much of the results expected for ECO perturbations, but has a cleaner numerical behavior which allows us to more easily discern physically relevant results. After this analysis, we proceed to present the results for the dynamics of perturbations propagating on a Schwarzschild solution with a boundary condition which represents an ECO.

To summarize briefly, we find that the spectral instability present in the BH case \cite{Jaramillo2020} is also present for ECOs. Furthermore, despite the pseudospectrum contour lines that we find protruding into the unstable half of the imaginary plane, our numerical results provide strong indications that the evolution operator $e^{iL\tau}$ is in fact contractive. This result is further supported by an analytical calculation of the numerical abscissa. This entails the  absence of a transient growth but, following from the same analysis, we argue that pseudo-resonances may indeed trigger instabilities when second order effects are taken into account. 

\subsection{Toy model: the P\"oschl-Teller potential}\label{sec:results_PT}
\begin{figure*}[t!]
	\centering
	\includegraphics[scale=1.03]{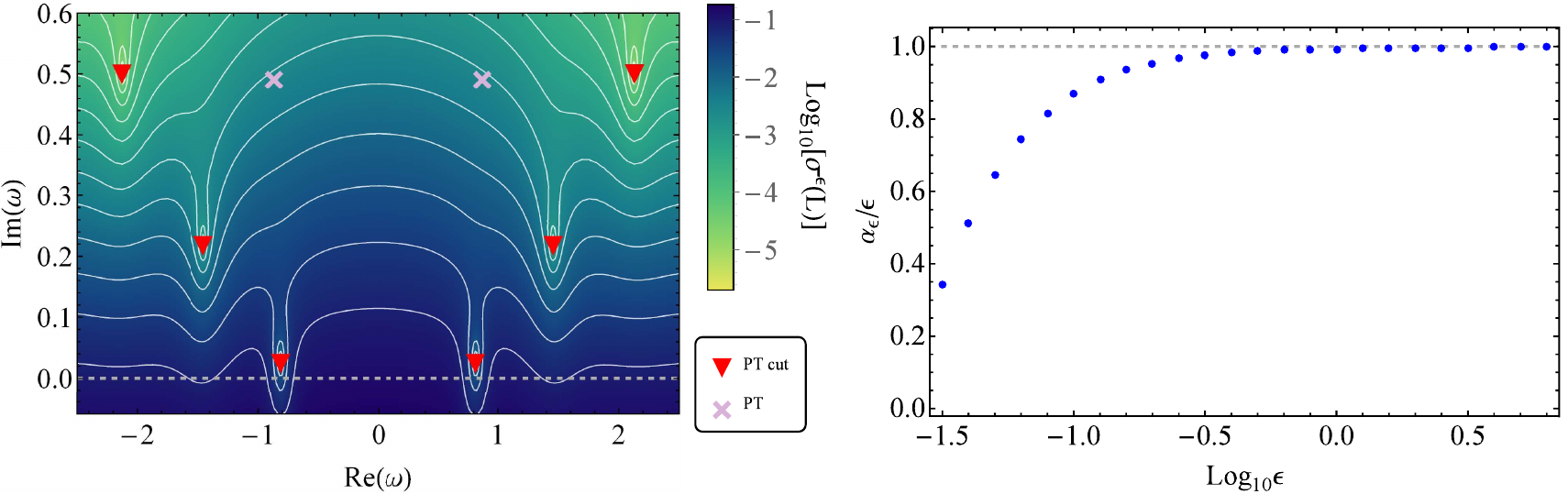}
	\caption{{\em Left Panel:} QNMs (red triangles) and pseudospectrum (white curves) of the $L$ operator for the PT potential, with a Dirichlet boundary condition at $\mathcal{E}=10^{-3}$ and $N=200$ spatial discretization points. Contour lines correspond to values between $-4.7$ (uppermost contour) and $-1.1$ (lowest contour) with steps of $0.3$ in the log scale. The horizontal dashed line indicates the position where $\text{Im}(\omega)=0$. Light-purple crosses indicate modes of the PT potential without the boundary condition. {\em Right Panel}: Lower bound $\alpha_\epsilon(L)/\epsilon$ to the evolution operator norm for different values of $\epsilon$, computed numerically from the pseudospectrum of the PT case with Dirichlet condition at $\mathcal{E}=10^{-3}$ and with $N=200$. The horizontal dashed lines corresponds to $\alpha_\epsilon(L)/\epsilon=1$.}
	\label{PTps}
\end{figure*}

The PT potential provides us with a clean setup for the pseudospectrum analysis. In the case of dissipative boundary conditions at both asymptotic regions (${\cal E}=0$), the QNMs are known exactly, and they follow a distribution in the complex plane very similar to the large frequency limit of BH QNMs. Additionally, the similarities between PT and BH potentials turn out to be not limited to their spectrum, as both models present a very similar pseudospectrum as well~\cite{Jaramillo2020}. The main reason for the PT potential's cleaner numerical behavior is its exponential asymptotic decay, which eliminates the branch cut present in the BH case. As discussed in ref.~\cite{Jaramillo2020}, this branch cut manifests numerically as non-convergent eigenvalues on the imaginary axis, which affect the behavior of the pseudospectrum in their vicinity.
 
The left panel of fig. \ref{PTps} shows the spectrum (red dots) and pseudospectrum (color code) for the PT-ECO with a Dirichlet boundary at $\mathcal{E}=10^{-3}$. We have used $N=200$ grid points in our discretization method, and have checked the robustness of the results by varying $N$ to larger values. The internal precision for all calculations is set to $\sim 200$ digits. For comparison, the figure also displays the original PT QNMs (in blue dots) obtained with dissipative boundary conditions, given by $\omega_n = \pm \sqrt{3}/2 + i \left(n+1/2\right)$. As opposed to the original PT QNMs, one observes that the fundamental mode of the PT-ECO resides very close to the real axis. Decreasing $\mathcal{E}$, this mode moves even closer to the real axis. This is an illustration of stable trapping and the appearance of long-lived modes, ubiquitously present in ECO configurations.

The proximity of QNMs to the real axis can be quantified in terms of the pseudospectrum: contour lines of $\epsilon\simeq 0.02$ and larger values cross this axis into the region of unstable modes (i.e. the negative imaginary half-plane, in our convention for Fourier decomposition). This is suggestive of a perturbation-triggered instability. According to eq. \eqref{pseudoDL}, one may presume that adding a perturbation $\delta L$ of minuscule $\|\delta L\|$ to the underlying operator could make the fundamental mode migrate into the unstable region. However, this migration is not guaranteed to occur in realistic physical scenarios. To understand this, take for instance the pseudospectrum of any self-adjoint operator, whose eigenvalues lie on the real axis. The $\epsilon-$contour lines are circles around these eigenvalues, and thus they necessarily spread out into parts of the complex plane with both positive and negative imaginary parts. While this does suggest that a perturbed version of such an operator could have modes growing boundlessly, it does not speak to the physicality of such a perturbation (i.e. this perturbation may need to change the problem structurally in a way which, for physical systems, does not occur naturally).

The most direct method of checking how this instability manifests itself is by explicitly adding  (following a systematic scheme) small perturbations $\delta L$ to the original operator and calculating the modified spectrum. Ref.~\cite{Jaramillo2020} explored this possibility by modifying solely the potential $\hat V$ within the operator $L_1$ in eq.~\eqref{eq:Operators_L1_L2}. In particular, {\em ad hoc} (but controlled and systematic) modifications of this potential in terms of random, or oscillatory perturbations were explored. Employing the same strategy here, we reached the same conclusions: the fundamental mode is not affected by ultraviolet perturbations to the potential, at least ones which preserve the asymptotic structure. To actually destabilize the fundamental mode and make it migrate to the unstable half of the complex plane with this type of small-scale perturbation turns out to require a more generic modification of the original operator as a whole. Changes in the fundamental mode are also expected when infrared effects are introduced, as noted originally in ref.~\cite{Nollert:1996rf} (see also discussions on infrared modifications in~\cite{Jaramillo2020,Qian:2020cnz,Liu:2021aqh}) and recently systematically addressed in ref.~\cite{Cheung:2021bol}. Analyzing whether any such perturbations can be related to a physically reasonable modification of our system is an intriguing question, but it falls beyond the scope of this work.

As discussed, the pseudospectrum also brings information about early time transients in the dynamical evolution under the unperturbed operator $L$. The notable protrusion of the pseudospectrum lines past the real axis is very suggestive for a transient period of growth for solutions to the homogeneous time-evolution problem \eqref{evolution_operator}. The pseudospectral abscissa \eqref{eq:pseudospectra_abscissa} quantifies the size for which such protrusion may impact the initial dynamics, in particular via the ratio $\alpha_\epsilon/\epsilon$. The right panel of fig.~\ref{PTps} presents the dependence of the ratio $\alpha_\epsilon/\epsilon$ on the parameter $\epsilon$. We observe a tendency toward a value $\sim 1$, which suggests a Kreiss constant $\mathcal{K}(L)=1$ achieved through \eqref{kreiss} at $\epsilon\to\infty$. This result strongly suggests that the evolution operator $e^{iL\tau}$ is in fact contractive, and therefore does not bring about any transient growth.

On the other hand, such protrusion of the pseudospectrum into the lower half-plane does indeed indicate the ease with which a pseudo-resonance can be triggered if the evolution equation had a time dependent external source, whose Fourier transform had a (real) frequency close to the pseudospectrum protrusion region, as discussed in \cite{Trefethen:1993,Jaramillo:2022kuv}. We will elaborate on the implications this may have in combination with the non-linearity of gravitational dynamics in the end of this section. First we proceed to present the analogous of the above results for perturbations on an ECO spacetime with an external Schwarzschild geometry.

\subsection{Pseudospectrum of ECOs}

\begin{figure}[t!]
	\centering
	\includegraphics[scale=1.04]{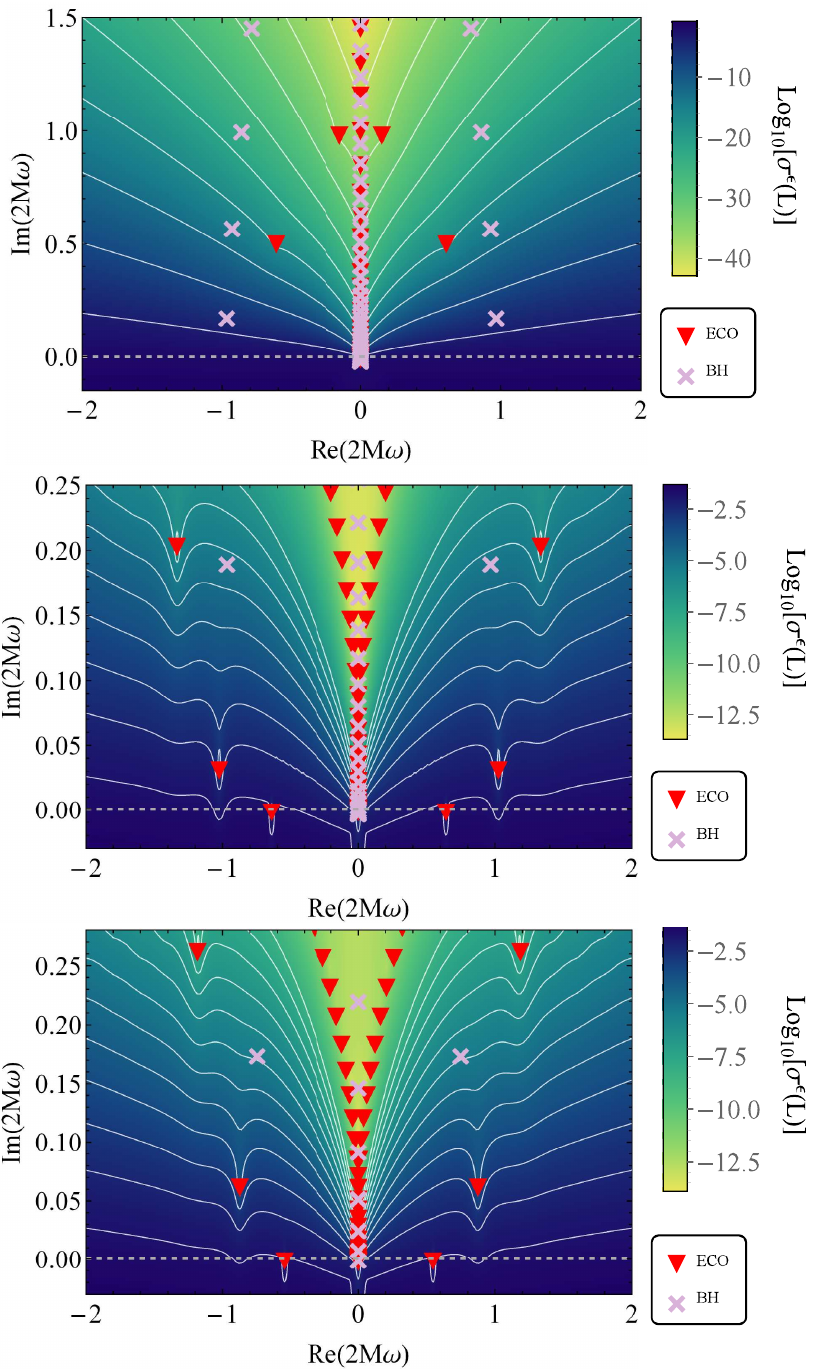}
	\caption{
	ECO's QNMs (red triangles) and pseudospectra for several configurations. The plots also display the respective BH QNMs (ligt-purple crosses), as well as non-convergent eigenvalues of the operator resulting from a discretised branch cut spreding from $\omega=0$.
	\emph{Top Panel}: Scalar perturbation for $\ell=2$ with ECO compactness $\mathcal{E}=1$, calculated with $N=200$ grid points. The contour lines range from $-44$ (uppermost contour) to $-3.5$ (lowest contour) with steps $4.5$ in the $\log$ scale.
	\emph{Middle Panel}: Same as top panel but with ECO compactness $\mathcal{E}=10^{-3}$. The contour lines range from $-6.5$ (uppermost contour) to $-1.7$ (lowest contour) with steps $0.6$ in the $\log$ scale. 
	\emph{Bottom Panel}: Same as middle panel, but for the axial modes of the gravitational field. For ECO's surfaces far from the horizon (top panel), the ECO QNMs lie far from the real axis. For smaller values of $\mathcal{E}$ (middle and bottom panels), the fundamental mode becomes long lived with a very small imaginary part. The pseudospectrum indicates and QNM instability for all cases.	
	}
	\label{figure:1}
\end{figure}

We now present the results for the pseudospectrum, and quantities derived therefrom, in the case of ECOs with a reflective surface of compactness dictated by the parameter $\mathcal{E}$. To ensure the validity of our results, we have first performed convergence tests for the QNM spectrum with an increasing number of grid points, and we have found that the value $N=200$ is sufficient for accuracy in the region of the complex plane we analyze (as long as $\mathcal{E}\gtrsim10^{-3}$). We have also compared the resulting QNMs with those obtained by a shooting method and found significant agreement. The internal precision for all calculations is set to $\sim 300$ digits. As opposed to the PT potential, the algebraic decay of the Regge-Wheeler potential towards infinity implies the existence of a branch cut, which in our notation is set to lie along the positive half of the imaginary axis. With our discretized operator, this branch cut yields non-converging (in the numerical resolution $N$) modes spreading out from the origin.

\subsubsection{QNM instability}
Figure \ref{figure:1} shows several pseudospectra which are representative for the ECO case, with varying compactness and different spin of the perturbed field. The top and middle panels compare the results for the $\ell=2$ mode of a scalar field ($s=0$) with two different compactness parameters ${\cal E}$. For less compact objects, such as ${\cal E}=1$ (top panel), the ECO QNMs (red dots) have a spectral abscissa (the imaginary part of the fundamental mode) which is larger than that of a BH with the same mass (blue squares), though the order of magnitude stays approximately the same. As the object becomes more compact ($\mathcal{E}=10^{-3}$, middle panel), the spectral abscissa can decrease by orders of magnitude, i.e. the fundamental QNM of ECOs can become extremely long-lived (for the $\mathcal{E}=10^{-3}$ case the spectral abscissa becomes of the order $10^{-4}$ in units of the Schwarzschild radius). The exact same behavior is found for gravitational ($s=2$) perturbations as well (bottom panel) \cite{Cardoso:2016rao}. The mechanism behind this is attributed to stable trapping of perturbations between the surface of the ECO and the photon sphere and leads to the formation of quasibound states \cite{Dolan:2007mj,Rosa:2011my,Cardoso:2011xi,Pani:2012bp,Destounis:2019hca,Vieira:2021xqw,Vieira:2021ozg} that repeatedly excite the photon sphere modes, giving rise to echoes in the ringdown signal \cite{Cardoso:2016oxy}, as well as an overall slower decay characterized by the actual ECO's QNMs.

In all these cases, the opening of the $\epsilon$-pseudospectrum contour lines across the complex plane is indicative of QNM instability, just as observed in the BH case~\cite{Jaramillo2020,Destounis:2021lum}, as well as in the PT-ECO toy model from the previous section. The main difference with respect to the BH case is precisely the closeness of the fundamental mode to the real axis. As it was the case in the PT-ECO model, for sufficiently small values of ${\cal E}$,  $\epsilon-$pseudospectral contour lines for small $\epsilon$ cross the real axis and protrude into the region of the complex plane where unstable QNMs would reside. In the examples of fig.~\ref{figure:1} the crossing happens for $\epsilon\gtrsim 10^{-3}$. Moreover, the protrusion is quite visibly not only for the long-lived fundamental QNM, but also for the first overtone, where it occurs for contour lines $\epsilon\gtrsim 10^{-2}$.

\begin{figure*}[t]
	\centering
	\includegraphics[scale=1.05]{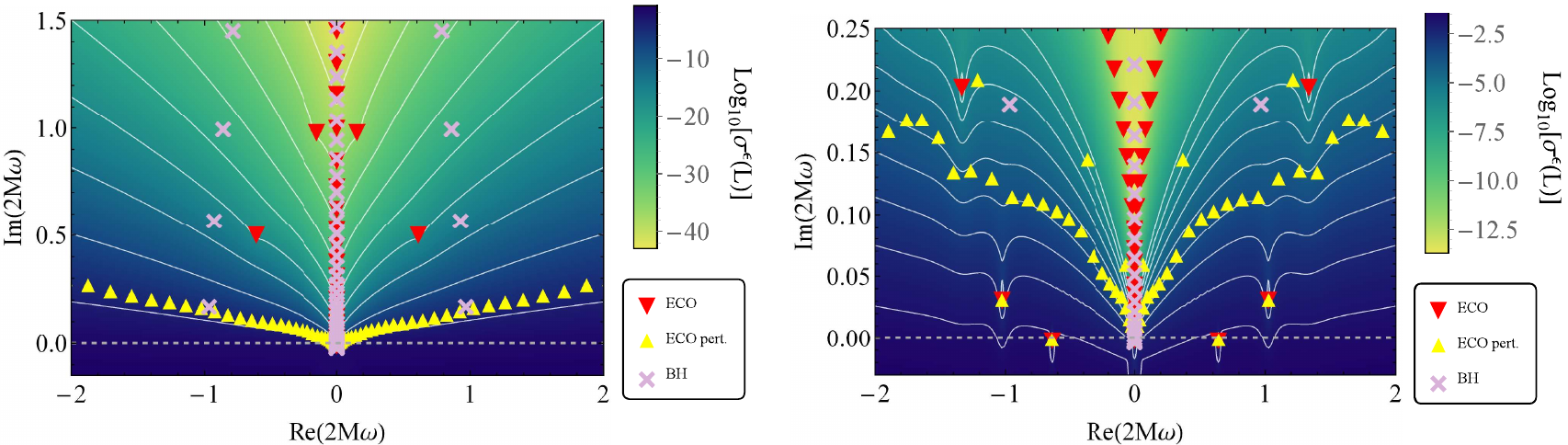}
	\caption{Top two panels of fig. \ref{figure:1} with the addition of the spectrum of $L$ with a perturbed version of the effective potential (upper-facing yellow triangles). The red triangles and light-purple crosses again designate the ECO and Schwarzschild BH QNMs, respectively. \textit{Left Panel}: ECO with compactness $\mathcal{E}=1$, and a random perturbation to the effective potential of energy norm $10^{-2}$. We observe that the fundamental mode is spectrally unstable.
		\textit{Right Panel}: ECO with compactness $\mathcal{E}=10^{-3}$, and a random perturbation to the effective potential of energy norm $10^{-1}$ (the largest reasonable perturbation). We observe that the fundamental mode and the first overtone are spectrally stable.}
	\label{figure:3a}
\end{figure*}

To better understand this instability, we have once again analyzed perturbations $\delta L$ (with $\|\delta L\|\simeq\epsilon$) which consist of adding a random or high-frequency sinusoidal function to the effective potential in $L_1$, analogously to the analysis performed in \cite{Jaramillo2020}. The fundamental QNM of ECOs appears to be stable all the way up to perturbations of energy norm $\epsilon\gtrsim 10^{-1}$ and higher, suggesting that a QNM migration to the unstable side of the complex plane is not triggered by ultraviolet effects. Akin to the discussion in the PT case, a possible destabilization of the fundamental mode into the unstable region requires a more generic modification on structure of the discretized $L$ operator, and a general disturbance of any kind to its position requires at least the inclusion of infrared effects~\cite{Jaramillo2020,Cheung:2021bol}.

When the ECO is less compact and the reflective surface reaches the photon sphere (e.g. top panel of fig. \ref{figure:1}, where $\mathcal{E}=1$), then we observe a different response to perturbations. Particularly, when the ECO's fundamental mode acquires a larger imaginary part, the $\epsilon$-pseudospectrum contour lines no longer protrude into the unstable part of the complex plane for small values of $\epsilon$, though their opening in the complex plane still indicates QNM instability. Interestingly, the fundamental mode for these configurations are susceptible to destabilization by an ultraviolet perturbation in the effective potential. This appears to be the case when the spectral abscissa is larger than that of the Schwarzschild BH (of the same mass). In other words, the stability of the fundamental mode against these types of operator perturbations (with energy norm up to $10^{-1}$) is present only when the spectral abscissa is of the same (or smaller) order of a Schwarzschild BH.

As evidence for this, fig.~\ref{figure:3a} shows the spectrum of the $L$ operator with a perturbation added to the effective potential which maximizes the frequency that can be resolved with the particular resolution used. The perturbation physically translates to a set of random numbers added to the potential which simulate very high frequency sinusoidal perturbations and solely depend on the energy norm of the perturbation \cite{Jaramillo2020}. For the first two cases of fig.~\ref{figure:1} ($\ell=2$ component of a scalar perturbation on $\mathcal{E}=1$ and $\mathcal{E}=10^{-3}$ ECOs) we perform a QNM calculation of the perturbed potential and find the following: in the first case, the fundamental mode is clearly unstable under a rather small perturbation of energy norm $10^{-2}$, while in the second case, even perturbations of up to $10^{-1}$ are insufficient to disturb the fundamental mode and even the first overtone. We also observe that, much like in the BH case~\cite{Jaramillo2020}, the branches of the perturbed modes approximately follow the pseudospectral contour lines.

\begin{figure*}[t]
	\centering
	\includegraphics[scale=1.05]{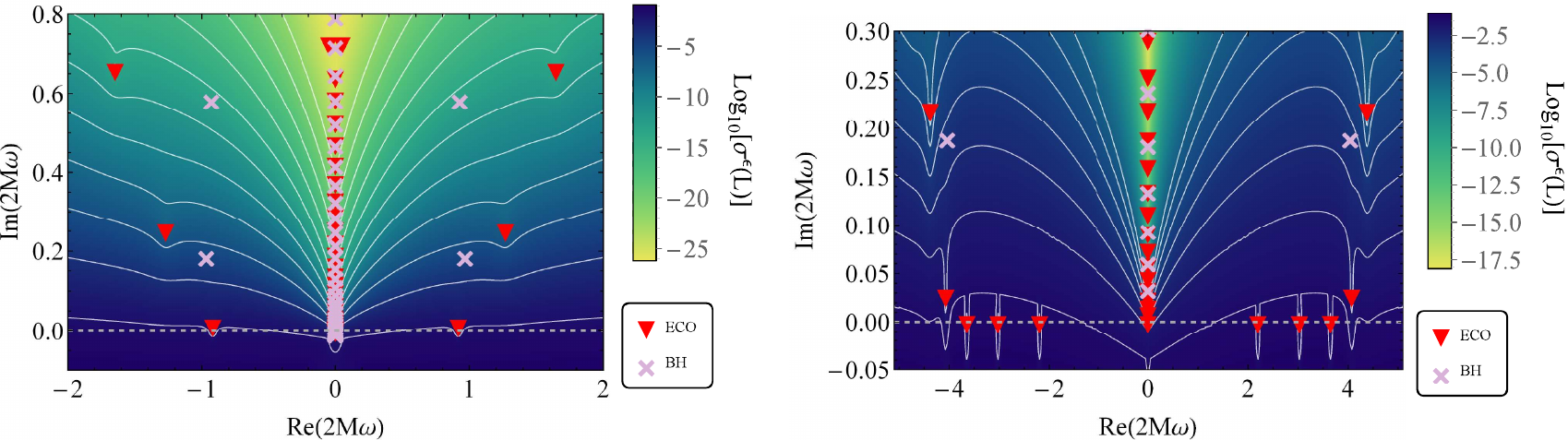}
	\caption{
		{\em Left Panel:} $\ell=2$ scalar QNMs (red triangles) and pseudospectra (white curves) of an ECO with $\mathcal{E}=10^{-2}$, calculated with $N=200$ grid points. The contour lines range from $-21.7$ (uppermost contour) to $-1.7$ (lowest contour) with steps $2$ in the $\log$ scale. For reference, we include the $\ell=2$ scalar QNMs of a Schwarzschild BH (light-purple crosses).
		{\em Right Panel:} $\ell=10$ scalar QNMs (red triangles) and pseudospectra (white curves) of an ECO with $\mathcal{E}=10^{-2}$, calculated with $N=200$ grid points. The contour lines range from $-6.5$ (uppermost contour) to $-1.7$ (lowest contour) with steps $0.6$ in the $\log$ scale. We again include the $\ell=10$ scalar QNMs of a Schwarzschild BH (light-purple crosses).
	}
	\label{figure:1a}
\end{figure*}

We complete the discussion by comparing the ECOs QNMs and pseudospectra for different angular $\ell$-modes. The key motivation behind this comparison is the argument that ECO structural instability is triggered in regime of high angular modes $\ell \gg 1$ \cite{Keir:2014oka,Cardoso:2014sna}. For numerical convergence reasons, we move up to the case of a slightly less compact ECO, with ${\cal E}=10^{-2}$. The left panel of fig. \ref{figure:1a} shows results for the scalar field with $\ell=2$, where the result is much the same as for the $\mathcal{E}=10^{-3}$ case, though the spectral abscissa is slightly larger. However, this is in strong contrast to the case $\ell=10$ (right panel of fig. \ref{figure:1a}), where modes become substantially longer-lived, leading to a plethora of protrusions of pseudospectrum contours into the unstable regime. In particular, we observe this for the first four modes (the fundamental mode and first three overtones) in the figure.

To better quantify how these protrusions into the unstable part of the complex plane affects the field's dynamics, we now once again proceed to apply the tools introduced in sec.~\ref{sec:Transient}.

\subsubsection{Transient effects}
As discussed above and in further detail in~\cite{Jaramillo:2022kuv}, the multifaceted nature of pseudospectrum analysis allows one to extend the study beyond the mere spectral instability effects, and obtain information on the early (and intermediate) time evolution governed by \eqref{matrix evolution}. Repeating the calculation performed for the PT example in sec.~\ref{sec:results_PT}, we calculate the ratio $\alpha_\epsilon/\epsilon$ appearing on the right-hand side of eq.~\eqref{bound}, which quantitatively characterizes the effects of the $\epsilon-$pseudospectral protrusion to the unstable half of the complex plane. 

\begin{figure}[t]
	\centering
	\includegraphics[scale=1.04]{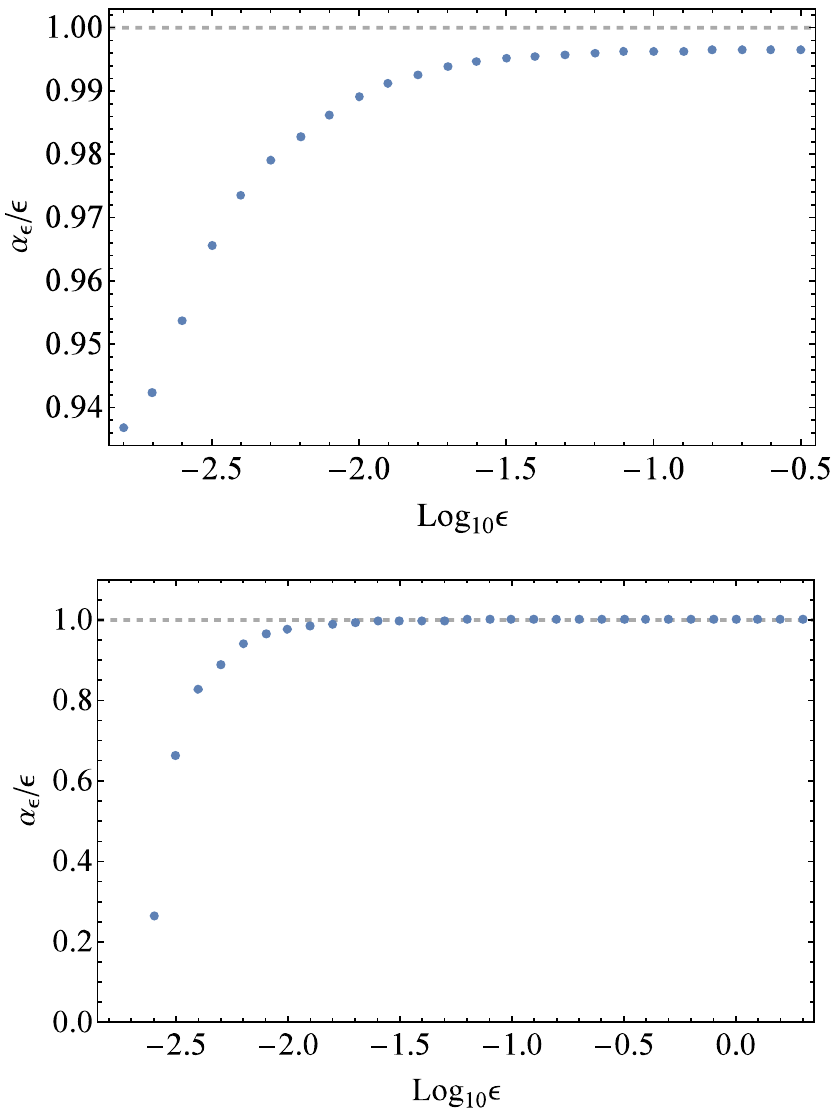}
	\caption{Lower bound $\alpha_\epsilon(L)/\epsilon$ to the evolution operator norm for different values of $\epsilon$.
	\emph{Top Panel}: scalar field with the $\ell=2$ with ECO compactness $\mathcal{E}=10^{-3}$. 
	\emph{Bottom Panel}: scalar field with the $\ell=10$ with ECO compactness $\mathcal{E}=10^{-2}$.
	Both cases indicate that $\alpha_\epsilon(L)/\epsilon\rightarrow1$ (horizontal dashed line showing the limit), implying a Kreiss constant $\mathcal{K}=1$ reached at $\epsilon\to\infty$, which occurs when the evolution operator is $e^{iL\tau}$ contractive.}
	\label{fECObound}
\end{figure}

As representative examples, fig.~\ref{fECObound} shows the results obtained from the cases with largest protrusion of pseudospectral lines for the scalar field: with parameters ${\cal E}=10^{-3}$, ${\ell=2}$ (top panel) and ${\cal E}=10^{-2}$, ${\ell=10}$ (bottom panel); the calculation in other cases leads to the same qualitative result. The plots provide strong indications that the ratio $\alpha_{\epsilon}/{\epsilon}$ tends asymptotically to unity from below as $\epsilon\rightarrow \infty$. This result agrees with the one observed in the PT-ECO toy model, and indicates that the Kreiss constant appears to attain the value ${\cal K}(L) = 1$ through this ratio at $\epsilon\to\infty$, which is consistent with the time evolution operator $e^{iL\tau}$ being contractive. This is indeed confirmed by the analytical calculation of the numerical abscissa presented below. The results for gravitational perturbations are qualitatively the same.

It is also worth noting that, as discussed in \cite{Jaramillo:2022kuv}, the pseudospectrum at large values of $\epsilon$  contains information about the early-time evolution of the system, as can be seen by the fact that the $\epsilon\to\infty$ limit constrains initial growth. The similarities between the pseudospectra of BHs and ECOs in this limit (which can be seen by the analogous behavior for the numerical and pseudospectral abscissa between the two cases) may well be what encodes the similarity between the early-stage ringdown of the two objects, which has thus far been a mystery form the point of view of spectral analyses. This phenomenon is further discussed in point 3.5. of ref.~\cite{Jaramillo:2022kuv}.

\subsection{Contractive operator and pseudo-resonances}

Our numerical results strongly suggests that the ECO evolution operator $e^{i L \tau}$ is contractive, i.e. $\|e^{iL\tau}\|\leq 1, \forall \tau>0$ (more specifically, the unity Kreiss constant ${\cal K}(L) = 1$ that we find is a necessarily condition for contractibility, actually strongly supporting the latter), regardless the potential dictating the dynamics (PT or Regge-Wheeler with different values of $\ell$). Here, we prove this result analytically by evaluating the
numerical abscissa and confirming its vanishing $\omega(L)=0$, that actually implies the contractive character of $L$~\cite{Trefethen:2005,Jaramillo:2022kuv}. We also argue that the $\epsilon-$pseudospectral protrusions into the unstable half of the complex plane are indicative of pseudo-resonances on second-order perturbations, possibly triggered by first-order perturbations and which can themselves trigger resonant-like instabilities breaking down the perturbative analysis.

\subsubsection{Numerical abscissa: an analytical result}

Following the procedure laid out in~\cite{Jaramillo:2022kuv} we use the formal expression \eqref{eq:BoundaryOp} of the boundary operator $L^\partial$, which determines the adjoint $L^\dagger$, to analytically calculate the numerical abscissa $\omega(L)$. In accordance with eq.~\eqref{eq:numericalabsicissa_eignvalue}, we obtain $\omega(L)$ from the largest eigenvalue of the operator
\begin{eqnarray}
\label{eq:Op_NumAbscissa}
\dfrac{1}{2i}\left(L^\dagger - L\right) &=& \frac{L^\partial}{2i} \nn \\ 
&=& \begin{pmatrix}
0 & 0\\ 0 & -\gamma  \delta(\chi - \chi_{_{\scri^+}})
\end{pmatrix}.
\end{eqnarray}
From eqs.~\eqref{eq:gamma_PT} and \eqref{eq:gamma_Schwarzschild}, we obtain $\gamma(\chi_{_{\scri^+}})=1$ in both the scenarios modelled by the PT and Regge-Wheeler potentials. In fact, we expect this result to be valid for any potential, since $\gamma(\chi_{_{\scri^+}})$ is determined directly from its asymptotic decay, i.e. the asymptotic flatness of the spacetime in the ECO case.

The eigenvalues of this diagonal operator~\eqref{eq:Op_NumAbscissa} are given by its diagonal entries, which either vanish or are negative. Since the numerical abscissa corresponds to the maximum eigenvalue, we then conclude
\begin{equation}
\omega(L) = 0.
\end{equation}
This result also implies ${\cal K}(L)=1$, as commented above, and as confirmed numerically. It allows us to conclude that the evolution operator is indeed contractive and exhibits no transient growth, in complete analogy to the BH case~\cite{Jaramillo:2022kuv}.

We emphasize that this result is a direct consequence of the fully reflective boundary condition imposed at the ECO surface via eq.~\eqref{BC_FirstOrderRed}, which eliminates part of the boundary terms which would appear in \eqref{eq:BoundaryOp}. Whether this is still the case for more generic conditions at the ECO surface is an open question. 

\subsubsection{Pseudo-resonances and non-linear bootstrap instability}
The pseudospectrum is a tool which describes the evolution of perturbations on BH and ECO spacetimes in more detail than a simple modal analysis can, in the present context of the non-selfadjoint character of the time evolution generator $L$. However, the evolution it pertains to is still only at the linear level. Even though in hydrodynamics the presence of linear-level transient effects have been used to infer the triggering of non-linear instabilities~\cite{Trefethen:1993}, in the ECO systems we analyzed here no such transient effects have been observed. This analysis therefore appears insufficient to capture the expected non-linear dynamical instabilities discussed in refs.~\cite{Cardoso:2014sna,Keir:2014oka}.

That said, the pseudospectrum of these systems turns out to be useful even beyond linear order. In particular, the transitive protrusion of pseudospectral lines through the real axis, along with the notion of pseudo-resonance ultimately associated with the third pseudospectrum definition~\eqref{eq:pseudospectra_def3}, can be combined with GR perturbation theory at second order~\cite{Cunningham1980,Gleiser:1996yc,Miller:2016hjv} to give rise to a potential instability mechanism. Such a mechanism is described in appendix A of ref.~\cite{Jaramillo:2022kuv}. Although that mechanism does not prove useful in the context of~\cite{Jaramillo:2022kuv}, it seems perfectly suited to the current ultra-compact object setting. We follow appendix A in~\cite{Jaramillo:2022kuv} in the discussion below.

Due to the non-linearity of GR, the dynamics of a perturbation at second order $u^{(2)}$ is always sourced by the first order field $u^{(1)}$ and, in particular gauges, (the homogeneous part of) the second order evolution operator is the same as the first order one. Thus, we can extend the linear system eq.~\eqref{matrix evolution} to second order as~\cite{Jaramillo:2022kuv}
\begin{eqnarray}
\left(\partial_\tau - i L \right)u^{(1)} &=& 0  \\
\label{eq:secondorder_system}
\left(\partial_\tau - i L \right)u^{(2)} &=& S[u^{(1)}].
\end{eqnarray}
Because the field $u^{(1)}$ does not undergo any transient growth in its evolution, we follow \cite{Trefethen:1993,Schmid:2007,Jaramillo:2022kuv} to model the source as $S\sim e^{i\omega\tau}s(\chi)$. Then, a formal solution to eq.~\eqref{eq:secondorder_system} follows by acting with the Green function on the source term via
\begin{eqnarray}
u^{(2)}(\tau,\chi) &=& i e^{i\omega \tau}\left( L - \omega {\mathbb I}\right) ^{-1} s(\chi) \nn \\
&=& i e^{i\omega \tau} R_{L}(\omega) s(\chi).
\end{eqnarray}
Note that in the second line we have used the fact that  (the integral kernel of) the resolvent $R_{L}(\omega)$ defined in eq.~\eqref{pseudoDLg} is precisely the Green function of the system. Thus, an upper bound estimate for the response in the solution $u^{(2)}$ to the oscillating source term reads
\begin{eqnarray}
\label{eq:secondorder_estimative}
\| u^{(2)}\| \lesssim e^{- {\rm Im}(\omega) \tau} \| R_{L}(\omega) \| \| s_n \|.
\end{eqnarray}
Remarkably, as a consequence of the perturbation theory structure in GR~\eqref{eq:secondorder_system}, the resolvent norm $\| R_{L}(\omega_n) \|$ which appears here is precisely the one used to define the pseudospectrum of the {\em linear operator at first order}, precisely the one we have studied in detail in previous sections. 

For standard resonances, the field's maximal response will be at the QNM spectral frequencies $\omega_n$, since $\| R_{L}(\omega_n) \| \rightarrow \infty$. However, in the non-selfadjoint case and due to the spread of $\epsilon$-pseudospectra sets
in the complex plane, we may also encounter resonant-like behaviour, namely pseudo-resonances, for which the resolvent's norm is still large, though not strictly divergent. Such pseudo-resonances are present in every pseudospectrum of a non-selfadjoint operator. But what makes the pseudospectrum of ECOs to stand out is the fact that for very compact objects ($\mathcal{E}\ll 1$) there exist strong protrusions in the unstable part of the complex plane {\em crossing} the real line. As a consequence, the norm of the resolvent $R_L(\omega)$ becomes very large for certain values along the real line. This is the key feature for the pseudo-resonant mechanism. Specifically, considering a time-depending external source to $u^{(2)}$ (here provided by $u^{(1)}$), through its Fourier transform, it can be seen as a superposition of {\em real frequency} harmonic contributions. As a consequence, if the Fourier-transformed external source is peaked at the (real) frequencies where the pseudospectrum protrudes into the unstable complex half-plane, a pseudo-resonance occurs. Quantitatively, the maximal response  growth-factor $R_{\mathrm{max}}$ that can be attained~\cite{Jaramillo:2022kuv} is given by
\begin{equation}
\label{e:Rmax_real}
\begin{split}
R_{\mathrm{max}} =\sup_{\omega\in\mathbb{R}} R_{\mathrm{max}}(\omega)& =
\sup_{\omega\in\mathbb{R}}\|(\omega - L)^{-1}\| \\& = \sup_{\omega\in\mathbb{R}}\|R_L(\omega)\|.
\end{split}
\end{equation}

This pseudo-resonance mechanism may well lead to a growth of $u^{(2)}$, sourced by $u^{(1)}$, which breaks down the perturbative expansion and leads to the non-linear dynamical instabilities referred to in~\cite{Cardoso:2014sna,Keir:2014oka,Cunha:2022gde}. This is in analogy to what occurs in hydrodynamics~\cite{Trefethen:1993}, though without the transient growth: whereas in the hydrodynamical case it is the non-selfadjoint linear transient growth that triggers the non-linear response in a bootstrap mechanism, in the gravitational case the bootstrap aspect is provided by the GR perturbative structure that source gravitational second-order perturbations with gravitational first-order ones.

\section{Discussion}\label{sec:Conclusion}
The pseudospectrum analysis offers a profound insight into the evolution dictated by linear operators and it has been widely applied in physical systems~\cite{Trefethen:1993,Driscoll:1996,Trefethen:2005,Schmid:2007}. Its ability to capture not just the poles, but the whole `topographical structure' of the Green's function provides a powerful tool to assess underlying non-modal effects, such as spectral instabilities, isospectrality properties, early-time transient phenomena, turbulence and pseudo-resonances --- features that an eigenvalue analysis of the system's operator does not account for. Only recently has the pseudospectrum analysis been introduced to the BH literature \cite{Jaramillo2020,Jaramillo:2021tmt,Destounis:2021lum,Gasperin2021,Jaramillo:2022kuv}. The pseudospectrum tool was first used to assess BH QNM spectral instabilities in Schwarzschild~\cite{Jaramillo2020,Jaramillo:2021tmt} and Reissner-Nordstr\"om~\cite{Destounis:2021lum} BHs. Such instabilities may offer new challenges to GW data analysis \cite{Jaramillo:2021tmt,Gasperin2021}. Going beyond the study of spectral instabilities, ref.~\cite{Jaramillo:2022kuv} has shown the wide variety of applications of the pseudospectrum analysis in  gravitational settings, such as the study of binary BH systems by means of an effective (non-selfadjoint) linear description.

This work studies horizonless ECOs with the tools provided by pseudospectrum analysis. Our goal was twofold. First, expanding on the results from ref.~\cite{Jaramillo2020,Jaramillo:2021tmt,Destounis:2021lum,Gasperin2021} and searching for signatures of QNMs instabilities. Second, applying the formalism laid out in ref.~\cite{Jaramillo:2022kuv} to assess possible initial transients growths that could trigger the non-linear dynamical instability identified in~\cite{Cardoso:2014sna,Keir:2014oka}. 

To this end, we have followed refs.~\cite{Price:2017cjr,Jaramillo2020} and initially considered a toy-model built from the PT potential. Despite its simplicity, this model actually shares a lot of the qualitative features of an ECO case, while also providing a useful testing ground for our numerical method.

We found that ECOs with any compactness suffer from the same ultraviolet spectral instability in the overtones as BHs do. Additionally, we observed new aspects concerning the stability of the fundamental mode under ultraviolet modifications in the potential. 

A well-known phenomenon is that the ECO QNMs $\omega_n^{\rm ECO}$ differ significantly from those of a BH with the same mass $\omega_n^{\rm BH}$. For instance, the fundamental mode $\omega^{\rm ECO}_0$ for larger, less compact ECOs may acquire values with an imaginary part (spectral abscissa) larger than in the BH case. As the ECO becomes more compact, $\omega^{\rm ECO}_0$, as well as the overtones, get closer to the real axis, becoming long-lived as a consequence of the trapping which occurs between the ECO surface and the light-ring potential barrier \cite{Cardoso:2016rao,Cardoso:2016oxy}. We observe that the ECO fundamental mode is unstable under ultraviolet modifications of the potential only when ${\rm Im}(\omega^{\rm ECO}_0)>{\rm Im}(\omega^{\rm BH}_0)$. As $\omega^{\rm ECO}_0$ goes down towards the real axis, it becomes ultraviolet stable approximately when ${\rm Im}(\omega^{\rm ECO}_0)\simeq {\rm Im}(\omega^{\rm BH}_0)$. From there on, destabilizing the fundamental mode requires either modifications in the whole operator (not just the potential), or the introduction of infrared effects in the potential~\cite{Jaramillo2020,Cheung:2021bol}.

This result is of particular importance for ultracompact ECOs, whose fundamental mode lies very close to the real axis. For these objects, the $\epsilon$-pseudospectrum contour-lines  cross the real axis into the unstable region of the complex plane for relatively small values of $\epsilon$. In principle, the fundamental mode of a slightly perturbed version of the operator could then cross into the unstable half of the complex plane. However, this is not observed for any perturbation to the effective potential. The perturbing operator which could cause this migration would therefore likely have to substantially change the structure of the underlying problem (e.g. introducing new derivative terms, albeit with small multiplicative coefficients), which is not directly relatable to physical sources of perturbation.

At this point we again emphasize the difference between spectral instabilities and full dynamical instabilities. The former are observed in both BHs and ECOs, and lead to a significant migration of modes under small modifications in the scattering operator, but not necessarily to a growth in the norm of the time-evolving field (unless the perturbed modes enter the unstable region). The latter, characterized by a dynamical growth of the field, can be identified at the linear level either if modes assume values in the unstable region of the complex plane, or if non-modal effects induce an initial transient growth or a pseudo-resonant behavior. Although clearly different, the spectral and dynamical instabilities occurring in a non-selfadjoint setting are ultimately related, as is made apparent by the fact that they can both be characterized by the non-triviality of the pseudospectrum. In particular, a spectral instability is one of the indicators by which one typically predicts a dynamical instability through linear-order analysis, the other ones being the protrusion of the pseudospectral lines into the unstable part of the complex plane and transient growth. While the hydrodynamics analysis of ref.~\cite{Trefethen:1993} presents all three of these characteristics, the situation with ECOs is somewhat different.

Particularly, by following the steps of ref.~\cite{Jaramillo:2022kuv}, we have calculated the Kreiss constant and numerical abscissa and found no indication of transient growth. We conclude that the evolution operator $e^{iL\tau}$ is contractive, i.e. that the solutions to \eqref{evolution_operator} can only have a constant or decreasing norm. This result, in particular the analytical calculation of the numerical abscissa $\omega(L)=0$, turns out to strongly rely on the field boundary conditions we have chosen for the surface of the ECO~\eqref{BC_FirstOrderRed}. An interesting open question is what will occur when different boundary conditions are considered.

Nevertheless, an initial transient growth is not the only mechanism able to trigger a non-linear dynamical instability. The $\epsilon$-pseudospectrum protrusion into the unstable region of the complex plane shows the existence of pseudo-resonances at and below the real axis which, when appropriately excited, can potentially lead to a dynamical instability. In particular, due to the fact that the Green function for the second-order perturbation field is precisely given by the resolvent $R_L(\lambda)$ of the first-order operator (from which the pseudospectrum is obtained)~\cite{Jaramillo:2022kuv}, pseudo-resonant excitations may be observed at second order, with the first-order solution acting as a source. The second-order contribution to perturbations has been shown to be non-negligible (and potentially measurable) even in the case of a BH~\cite{Cheung2022,Mitman:2022qdl}, where long-lived modes and pseudo-resonances are generally absent. Here, we can estimate the amount by which second-order effects are amplified just through the pseudospectrum we have calculated. Our conclusion is that for ultracompact horizonless objects, for which the fundamental modes and first overtones lie close to the real axis, second-order perturbations will be amplified significantly---potentially enough to lead to the breakdown of linear perturbation theory, and consequently to the destabilization of the astrophysical object itself.

Overall, the pseudospectrum framework and non-modal analysis provide a large range of tools to study wave phenomena at linear level for non-conservative systems driven by non-selfadjoint time evolution operators, beyond standard (self-adjoint) modal analysis~\cite{Jaramillo:2022kuv}. Potential future works in this direction include introducing this methodology to a wider range of BH and ECO setups, with different asymptotic structures and symmetries (see e.g. \cite{Cardoso:2017soq,Cardoso:2018nvb,Destounis:2019hca,Destounis:2020pjk,Destounis:2020yav,Mascher:2022pku,Destounis:2022rpk,Yang:2022wlm}).

\begin{acknowledgments}
The authors thank Emanuele Berti for helpful discussions.
V.B. is funded by the Spanish Government fellowship FPU17/04471.
R.P.M acknowledges financial support provided by COST Action CA16104 via the Short Term Scientific Mission grant and from STFC via grant number ST/V000551/1.
V.C.\ is a Villum Investigator and a DNRF Chair, supported by VILLUM FONDEN (grant no.~37766) and by the Danish Research Foundation. V. B. and V.C.\ acknowledge financial support provided under the European
Union's H2020 ERC Advanced Grant ``Black holes: gravitational engines of discovery'' grant agreement
no.\ Gravitas–101052587.
This project has received funding from the European Union's Horizon 2020 research and innovation programme under the Marie Sklodowska-Curie grant agreement No 101007855.
We acknowledge financial support provided by FCT/Portugal through grants 
2022.01324.PTDC, PTDC/FIS-AST/7002/2020, UIDB/00099/2020 and UIDB/04459/2020.
This research project was conducted using the computational resources of ``Baltasar Sete-Sois'' cluster at Instituto Superior T\'ecnico.
This work was supported by the French ``Investissements d'Avenir'' program through
 project ISITE-BFC (ANR-15-IDEX-03), the ANR ``Quantum Fields interacting with Geometry'' (QFG) project
 (ANR-20-CE40-0018-02),  the EIPHI Graduate School (ANR-17-EURE-0002) and  
  the Spanish FIS2017-86497-C2-1 project (with FEDER contribution).
\end{acknowledgments}

\bibliography{Bibliografia}

\end{document}